\documentclass[10pt,twocolumn,aps,pre,superscriptaddress]{revtex4-2}

\usepackage{amsmath}
\usepackage{graphicx}
\usepackage{color}
\usepackage{upgreek}
\usepackage[linkcolor = blue, citecolor = blue, urlcolor = blue, colorlinks = true]{hyperref}
\usepackage[version=3]{mhchem}
\usepackage{commath,amssymb}
\usepackage{ushort}
\usepackage[dvipsnames]{xcolor}
\usepackage{tikz}
\usepackage{xfrac}
\usepackage{MnSymbol}
\usepackage{textcomp}

\graphicspath{{FIGURES/}}

\begin{document}

\author{Eren G{\"u}lmez}
\email{egulmez26@ku.edu.tr}
\affiliation{Department of Physics, Middle East Technical University, 06800 Ankara, T{\"u}rkiye}
\affiliation{Institute for Theoretical Physics, Utrecht University,\\ Princetonplein 5, 3584 CC Utrecht, The Netherlands}

\author{Joost de Graaf}
\email{j.degraaf@uu.nl}
\affiliation{Institute for Theoretical Physics, Utrecht University,\\ Princetonplein 5, 3584 CC Utrecht, The Netherlands}


\title{Growing Knot Embeddings into Spherical Graphs}
\date{\today}

\begin{abstract}
\noindent Knotted structures occur throughout soft matter and biology, where topological constraints can shape relaxation and spatial organization. Here, we introduce a computational construction that maps a specific 3D knot embedding to a graph embedded on the sphere through growth and interfacial-energy relaxation. For standard 3D embeddings of prime knots with at most seven crossings, we obtain a family of growth-based graphs that generally differ from the corresponding Tait graphs. We further identify an empirical growth-graph reconstruction rule based on polygonal regions with alternating crossing structure in a knot diagram. Non-minimal diagrams demonstrate that these graphs are not knot invariants, while selected connected sums suggest that graph features associated with their prime factors can remain recognizable. Determining how these graphs transform under isotopy may reveal invariant features, which we leave for future study.
\end{abstract}

\maketitle

\section{\label{sec:intro}Introduction}

\noindent Topological constraints play a central role in soft and biological matter~\cite{TUBIANA20241, Shankar_2022}. The embedding, entanglement, and connectivity of extended objects can alter accessible configurations, relaxation pathways, and collective organization~\cite{TUBIANA20241}. Knotted and linked structures occur in polymers~\cite{Micheletti_2011}, DNA~\cite{Wasserman_DNA}, filament networks~\cite{terriac2019vimentin}, vortex lines~\cite{KlecknerIrvine}, and confined elastic~\cite{Raymer2007} or active materials~\cite{Patil_2023}. In these settings, topology may influence transport, packing, mechanical response, and defect organization~\cite{BowickSymmetry, giomi2015geometrytopologyturbulenceactive}. This motivates interest in connecting the topology of an embedded object to spatial structures that arise through local interactions and relaxation.

Mathematically, a knot is an embedding of the circle $S^{1}$ in three-dimensional (3D) space, considered up to ambient isotopy~\cite{rolfsen2003knots}. A planar knot diagram records a generic projection together with over-under information at every crossing. Equivalent diagrams are related by planar isotopy and Reidemeister moves~\cite{LickorishW.B.Raymond1997Aitk}. Knot types can also be studied through algebraic invariants, including the Alexander, Jones, and HOMFLY-PT polynomials~\cite{Alexander1928TopologicalIO, Jones:1985dw, homfly, homflypt}. Such invariants are powerful but not complete,~\textit{i.e.}, distinct knots can share the same value of a given polynomial invariant~\cite{LickorishW.B.Raymond1997Aitk}.

Graph-theoretic constructions provide an intermediary between knot diagrams and knot invariants~\cite{Bollobas1998Modern_Ch10}. For example, given a knot diagram with a checkerboard shading, one obtains plane Tait graphs by assigning vertices to regions of one color and edges to crossings; crossing information may be retained through edge signs~\cite{KAUFFMAN1989105}. Knot invariants can be recovered by evaluating the polynomials of their associated graphs. For instance, Thistlethwaite's theorem expresses the Jones polynomial, up to normalization, as an evaluation of the ordinary Tutte polynomial of a Tait graph~\cite{THISTLETHWAITE1987297}. This was later extended to cover non-alternating knots~\cite{KAUFFMAN1989105} and embedded graphs~\cite{Bollobs2001API, Bollobs2002APO, Moffatt_2008}. Such a construction is typically associated with a chosen diagram rather than directly with a knot type. Changing the projection or checkerboard convention may change the graph, while knot invariance is recovered through normalized polynomial evaluations or equivalence under graph operations corresponding to Reidemeister moves~\cite{welsh_knotgraphs}.

Here, instead, we ask whether a graph can emerge directly from a 3D knot embedding. To tackle this problem, we take inspiration from biology and propose a growth-and-relaxation process that maps a knot to a graph. Conceptually, our approach aligns with ideas from geometric knot theory, in which spatial realizations are assigned to knot types by minimizing quantities such as rope length~\cite{Cantarella_2002}. Tight or ideal knots provide approximately canonical 3D representatives. In addition, their self-contact sets encode nonlocal geometric relations between distant portions of the knot~\cite{ashton2005selfcontactsets50tightly, Ashton_2011}. Our construction differs in that the knot is not tightened directly. Instead, growth drives the formation of a surrounding cellular contact network.

In brief, our growth approach is as follows, we represent a given knot embedding by a tubular neighborhood segmented longitudinally. The segments, which we call \textit{cells}, inherit a cyclic ordering from the knot. Each cell's volume grows at a constant rate, while an interfacial-energy algorithm relaxes the complete cellular partition. Our method is adapted from a level-set and auction-dynamics framework for cellular rearrangements in tissues by Mohammad~\textit{et al.}~\cite{mohammad2022numerical}. The growing union eventually becomes compact and approximately spherical. Fixing the initial tubular neighborhood and imposing connectivity preserve the segmented knotted scaffold throughout the growth process, which ensures that information about the knots spatial organization influences contacts at the outer boundary. That is, the spherical surface is partitioned by the cells, and contacts between cells that were not adjacent along the original knot can be used to define a surface network of cell borders.

We view this network as the skeleton of an effective tangent line field. Singular points of network valence $n$ are assigned index $q_{n} = 1 - n/2$, and each complementary surface domain is assigned one $+1$ index by a minimal-extension convention. For a connected cellular network on the sphere, this assignment is consistent with Euler's relation~\cite{Crowley2016ThePT}. Note that we use this exclusively as a geometric interpretation of the extracted network. The tangent-line field should not be interpreted as the outcome of an independently simulated orientational field. That is, our algorithm does \textit{not} solve a microscopic nematic free-energy model as,~\textit{e.g.}, in Refs.~\cite{nematic_defects, Bowick_2009}. The skeleton also represents a graph belonging to the embedded knot, which we will focus on in this paper.

To test our approach, we consider 3D embeddings of knots, whose projections realize standard minimal-crossing diagrams, for the fourteen nontrivial prime knots with at most seven crossings together with the unknot. Within this finite catalog, we obtain pairwise non-isomorphic growth-based graphs. These outputs generally differ from the corresponding Tait graphs. For two-strand torus knots, however, the growth-based graph agrees with one member of the checkerboard-dual Tait pair. We also identify an empirical reconstruction of the growth-based graph belonging to minimal knot diagrams in this catalog. This relies on placing vertices and associated edges in polygonal diagram regions with alternating over-under structure. We provide detailed rules for this reconstruction in the main text.

We observe that different non-minimal diagrams of the same knot can produce non-isomorphic graphs. Thus, the present construction does not identify a knot invariant. Applying our growth algorithm and empirical reconstruction technique to several selected sums knots, we nevertheless retain recognizable contributions from their prime factors. In addition, we recover elements of isotoped unknots in non-minimal sum-knot diagrams. These observations suggest that more systematically investigating how growth-based graphs change by isotopy is useful. This could lead to systematic understanding of the relation between the embedding the growth-based graph, and it may reveal new knot invariants. We leave this more involved task open for future studies.

The remainder of this paper is organized as follows. Section~\ref{sec:concept} introduces the segmented tubular representation and the surface-network construction. Section~\ref{sec:algorithm} specifies our algorithm, the way in which we extract graphs, and numerical choices. Section~\ref{sec:results} presents the prime-knot catalog, the empirical graph reconstruction from planar diagrams, selected connected sums, and non-minimal diagrams. Section~\ref{sec:discussion} contextualizes our work and discusses its limitations. Lastly, we close with a summary and brief outlook in Section~\ref{sec:close}.

\section{\label{sec:concept}Concept of the Construction}

\begin{figure}[!htb]
\includegraphics[width=0.5\columnwidth]{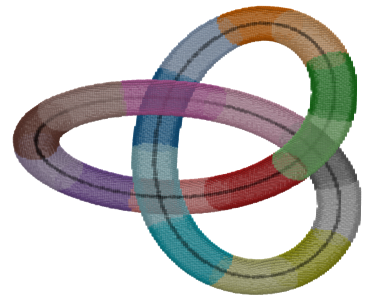}
\caption{\label{fig:tubular}\textbf{Segmented tubular representation of a trefoil.} The black curve is the knot centerline and the transparent tube is a tubular neighborhood used to initialize the simulation. Colors distinguish the cyclically ordered knot cells. The depicted initialization contains 20 knot cells.} 
\end{figure}

\noindent We first introduce the geometric construction before specifying details of the numerical algorithm. Let $K \subset \mathbb{R}^{3}$ be a smooth embedded representative of a knot type. A sufficiently thin tubular neighborhood of $K$ is a solid torus~\cite{rolfsen2003knots}. We partition this tube longitudinally into $N-1$ labeled regions, $C_{2}$, $\ldots$, $C_{N}$, whose labels follow the cyclic order of the knot. The exterior medium is denoted by $C_{1}$. Figure~\ref{fig:tubular} illustrates this representation for a trefoil.

\begin{figure*}[!htb]
\includegraphics[width=0.8\textwidth]{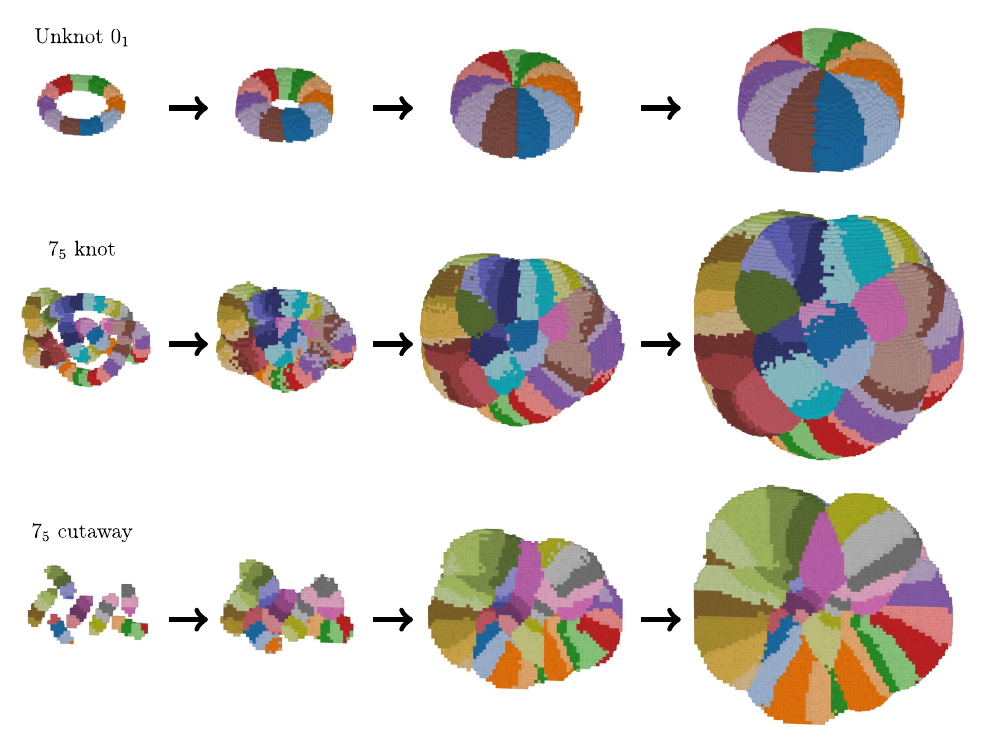}
\caption{\label{fig:cgrow}\textbf{Growth and relaxation of segmented knot embeddings.} Colors identify knot cells. The first and second rows show selected stages for the unknot and $7_{5}$ knot, respectively. The third row gives a cutaway view of the $7_{5}$ knot's trajectory, illustrating how cell regions remain connected to the tubular scaffold, while expanding toward the exterior boundary.} 
\end{figure*}

At each growth step, every knot cell increases its volume by a globally fixed increment. The complete system is then relaxed under weighted interfacial energy. Because every grid point carries exactly one label, cells do not overlap. Localized reassignment rules prevent cells from nucleating far from their existing region. We enforce that cells remain linked to the tubular region, which is immutable, and remain in contact with their neighbors. As the knot grows outward and the cell aggregate compacts, distant portions of the knot may come into contact. Figure~\ref{fig:cgrow} shows representative growth trajectories. Equal volume growth, interfacial relaxation, and an isotropic exterior favor compact shapes after sufficient growth.

\begin{figure}[!htb]
\includegraphics[width=1.0\columnwidth]{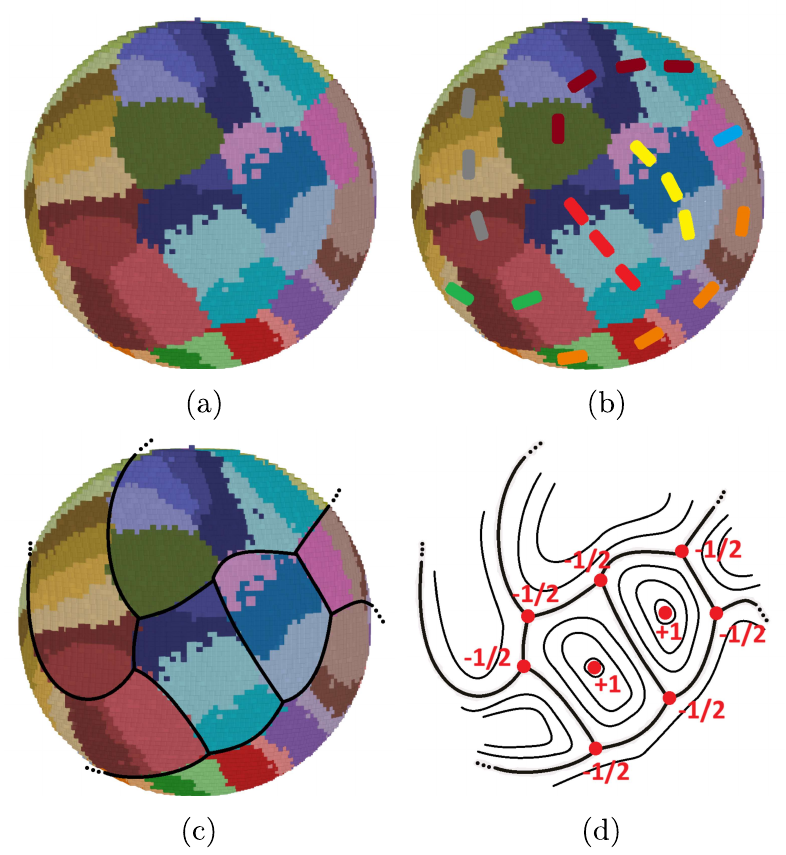}
\caption{\label{fig:cgraph}\textbf{Surface defects and spherical graph.} (a) Surface cell partition obtained from the grown $7_{5}$ knot embedding, see Fig.~\ref{fig:cgrow}. (b) Dashed connections indicate labels that were adjacent in the initial cyclic segmentation. (c) After merging through this adjacency relation, boundaries between nonadjacent cellular domains form the black network. (d) The network is interpreted as a line-field skeleton. A threefold junction has index $-1/2$; a domain interior is assigned one $+1$ index by the minimal-extension convention. The corresponding graph is embedded on the sphere and may be displayed in the plane \textit{via} stereographic projection.} 
\end{figure}

In the finite computations reported here, the outer aggregate boundary becomes sufficiently close to a sphere that the cells can be embedded onto $S^{2}$, see Fig.~\ref{fig:cgraph}(a). Cell labels that were consecutive along the initial knot are regarded as belonging to the same local strand. We therefore merge such surface patches and mark only contacts between nonconsecutive knot cells, see Fig.~\ref{fig:cgraph}(b,c). The resulting boundary network partitions the sphere into larger \textit{cellular domains}. This network can be interpreted as a separatrix skeleton of an effective tangent line field, see Fig.~\ref{fig:cgraph}(d). A network singularity of valence $n$ carries the index $q_{n} = 1 - n/2$. Thus, a terminal point has $q_{1} = +1/2$, a regular point has $q_{2} = 0$, a threefold junction has $q_{3} = -1/2$, and higher-order junctions carry increasingly negative index. Each complementary cellular domain is assigned one $+1$ singularity in the minimal extension. The unknot in its circular initialization constitutes a degenerate case, see Section~\ref{sec:results}. Unless stated otherwise, the line-field interpretation below is applied only to connected cellular networks. Lastly, the boundary network also defines a graph that can be stereographically projected onto the plane. This defines the objects (\textit{growth-based graphs}) of interest to our study.

\section{\label{sec:algorithm}Algorithm and Analysis}

\noindent To obtain growth-based graphs, we adapt the volume-constrained cellular-rearrangement algorithm of Mohammad~\textit{et al.}~\cite{mohammad2022numerical}. Let the computational domain $\Omega \subset \mathbb{R}^{3}$ be discretized on a regular grid (typically $70^3$ points), and let $\{C_{i}^{k}\}_{i=1}^{N}$ be the labeled partition at growth-relaxation step $k$; in our analysis we use $20 \le N \le 120$ and $20 \le k_{\text{max}} \le 35$. Interfaces are represented implicitly by label changes. For $i \neq j$, write $\Gamma_{ij} = \partial C_{i} \cap \partial C_{j}$ and assign a symmetric interfacial coefficient $\sigma_{ij} = \sigma_{ji}$ with $\sigma_{ii} = 0$; values are provided toward the end of this section. The sharp-interface energy is then written as
\begin{equation}
\label{eq:surface-energy} E(\{C_{i}\}) = \frac{1}{2} \sum_{i \neq j} \sigma_{ij} \operatorname{Area}(\Gamma_{ij}) ,
\end{equation}
where the factor $1/2$ prevents double counting.

For one relaxation step, the characteristic function $\chi_{C_{j}^{k}}$ of each label is convolved with a Gaussian heat kernel $G_{\Delta t}$, which depends on the time step $\Delta t$. The assignment cost for label $i$ is
\begin{equation}
\label{eq:assignment-field} \phi_{i}^{k}(x) = \sum_{j=1}^{N} \sigma_{ij} \left( G_{\Delta t} * \chi_{C_{j}^{k}} \right)(x).
\end{equation}
Thresholding these fields approximates multiphase curvature-driven motion~\cite{mbo, esedoglu_otto}. Auction dynamics~\cite{jacobs_merkurjev_esedoglu} imposes (discrete) target volumes, while localized bidding restricts a grid point to labels represented in its current neighborhood. Localization is implemented by a strong penalty term on costs $\phi_{i}^{k}(x)$ for regions outside of a boundary mask (generally with thickness $\delta_{\text{mask}} = 2$) defined for cell $i$. This mask suppresses nonlocal nucleation and cell splitting~\cite{mohammad2022numerical}.

We add to this basis that, at every growth step, the target volumes are updated by
\begin{align}
\label{eq:growth-knot} V_{i}^{k+1} &= V_{i}^{k} + g, && i = 2, \ldots, N ;\\
\label{eq:growth-background} V_{1}^{k+1} &= V_{1}^{k} - g(N-1) ,
\end{align}
where $i = 1$ labels the background as before and $g$ is an integer number of grid points. The latter shrinks by the amount the knot cells grow collectively, such that the total volume of the system is conserved. After the target update, we apply the localized auction dynamics with new target volumes. However, we impose three additional constraints: (i)~Each knot cell contains an immutable \textit{seed} region inherited from the original tubular neighborhood. (ii)~A candidate state is rejected if a knot-cell label becomes disconnected from its seed. (iii)~It is also rejected if seed-level cyclic adjacency is lost. These checks ensure that information about adjacency in seed region is carried over throughout the growth from the embedding to the outer surface of the emergent spherical cell aggregate.

After growth, we identify boundary grid points that separate $C_{1}$ from the knot-cell union. For a boundary point $x_{m}$ carrying knot label $i$, let the set $I_{i}$ be given by  $I_{i} = \{i-1, i, i+1\}$ with cyclic labeling over $2$, $\ldots$, $N$. Additionally, let $N_{m}$ be the set of knot labels in the $3 \times 3 \times 3$ neighborhood of $x_{m}$ after removing the background label. We mark $x_{m}$ as part of the nonlocal-contact network if $N_{m} \setminus I_{i} \neq \varnothing$. Connected marked points are thinned to a one-grid-point skeleton, components that connect to the rest of the network at only one endpoint containing fewer than $8$ grid points are removed. The remaining network defines a \textit{raw graph} embedded on $S^{2}$. To make this embedding explicit, the radial direction from the aggregate center to the surface is used to perform a projection. A \textit{cellular domain} is a maximal connected union of surface patches connected through labels that are consecutive along the knot.

Interfacial-energy models with isotropic tensions generically favor threefold junctions satisfying force balance, whereas higher-order junctions are structurally unstable and tend to split~\cite{plateau1873statique, taylor1976structure}. This behavior creates an ambiguity in the above graph construction. For example, a diagram region that suggests an ideal valence-$n$ junction with $n>3$ may appear after relaxation as a cluster of threefold junctions connected by short edges. The topology of the obtained network can then depend on small perturbations even when the large-scale contact pattern is unchanged. To address this issue, we perform a junction-coalescence regularization in our growth algorithm to obtain an \textit{idealized graph}.

Define $a(N_{m})$ as the number of adjacent label pairs in $N_{m}$. Provided $N_{m}$ does not contain every knot label, the number of cyclically connected label blocks is $r(N_{m}) = \vert N_{m} \vert - a(N_{m})$. Points with $r(N_{m}) \geq 3$ are candidate junction points for regularization. Maximal six-connected sets of candidate points define junction clusters $\mathcal{C}_{\alpha}$, with centroid $L_{\alpha}$ and attached label set
\begin{equation}
\widetilde{N}_{\alpha} = \bigcup_{x_{m} \in \mathcal{C}_{\alpha}} N_{m} .
\end{equation}
For every cluster $\alpha$ and $\beta$ separated by a distance less than $d_{\text{max}}$, labels present in one cluster but absent from the other receive an attractive modification of the assignment field centered at the opposite centroid,
\begin{equation}
\label{eq:attraction-field} \phi_{k}^{\mathrm{att}}(x) = \phi_{k}(x) - A \exp \left[-\frac{ \vert x - L_{\beta} \vert^{2} }{2 w^{2}} \right],
\end{equation}
with $k \in \widetilde{N}_{\alpha} \setminus \widetilde{N}_{\beta}$. The existing localization and the choice of $d_{\text{max}}$ ensures that the attraction does not reorganize the bulk scaffold. The amplitude $A$, width $w$, mask thickness, number of attraction iterations, and $d_{\text{max}}$ are numerical regularization parameters. A candidate state is rejected if new bridges between two separatrices are formed. Because junction-attraction algorithm becomes ineffective at small edge separations, we conventionally collapse edges having a distance less than $3$, which gives the idealized graph. An idealized graph is considered topologically stationary if it is unchanged for $5$ consecutive growth-relaxation steps, which we then identify as the graph of a knot embedding.

The dimensionless interfacial weights used throughout our calculations are
\begin{equation}
\label{eq:sigma-choice} \sigma_{ij} =
\begin{cases}
0, & i=j,\\
1.6, & \min(i,j) = 1,\ i\neq j,\\
0.8, & i,j\geq 2,\ i\neq j,\ i\sim j,\\
2.7, & i,j\geq 2,\ i\neq j,\ i\not\sim j.
\end{cases}
\end{equation}
where $i\sim j$ means that knot cells $i$ and $j$ are consecutive in the cyclic segmentation, including the pair $(2,N)$. The lower cost for consecutive knot cells favors continuity along the scaffold, whereas the larger cost for nonconsecutive cells discourages unnecessary contacts. The background-knot value controls compactification of the exterior interface and was given an intermediate value.

We use flexible growth and time step parameters such that the pair initialized at $(\Delta t_0, g_0) = (0.56, 28)$ are doubled whenever a knot-cell surpasses the volume thresholds of $V_i = 170$ and $V_i = 530$. These rules were found empirically to give faster convergence without compromising on smoothness of the growth process.  We typically start the junction-attraction algorithm once the final spherical surface is formed; however, for simulations where contractible junctions become widely separated during the growth, we use growth and junction-attraction algorithms simultaneously. We set junction-attraction parameters $w=5$ with $A$ varied within the range of $2.88\times 10^{-6}$ to $3.2\times 10^{-6}$. We set $d_{\text{max}}$ larger than any length in the grid, but if a simulation produces a candidate state that is rejected, then we repeat the simulation with $6 \le d_{\text{max}} \le 10$. We iteratively repeat the auction algorithm while dividing the initialization parameter $\varepsilon$ by $5$ at each step, from $10$ to $10^{-6}$. Our parameter choices, 3D embeddings, raw-graph outputs, and a robustness study are provided in the supplemental material accompanying this publication~\cite{ElectronicSupplement}.

\section{\label{sec:results}Results}

\noindent We first present the graph catalog for standard minimal-crossing embeddings of prime knots up to and including seven crossings. We then formulate an empirical reconstruction based on alternating polygonal parts of the considered minimal diagrams. We also examine selected connected sums and use non-minimal diagrams to probe the embedding-dependence of the idealized graphs.

\begin{figure*}[!htb]
\includegraphics[width=1\textwidth]{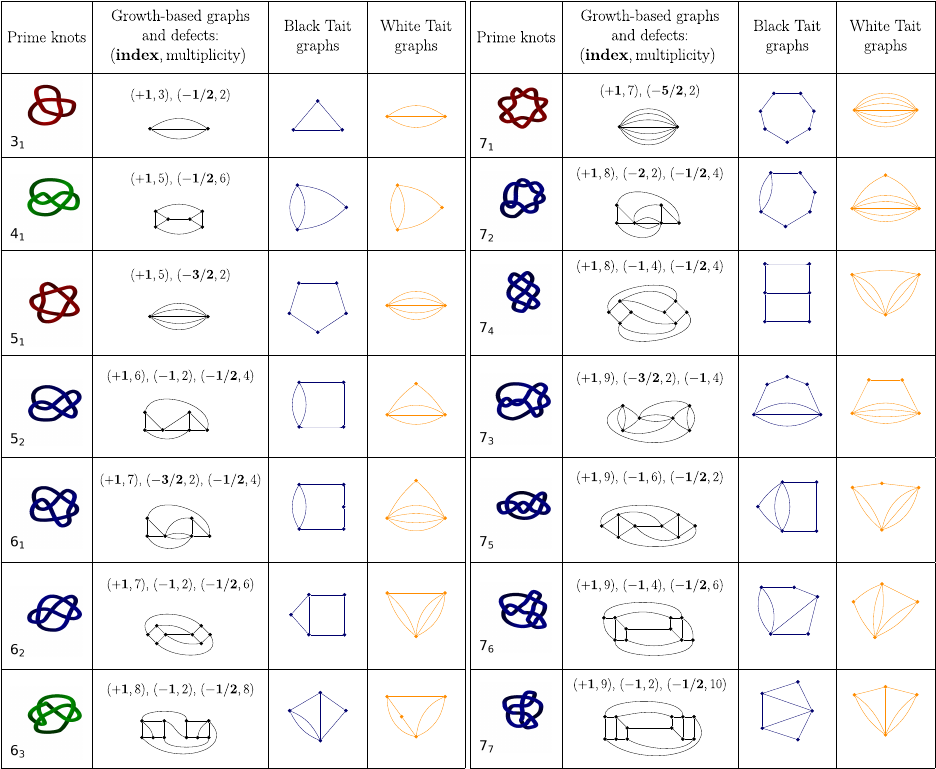}
\caption{\label{fig:prime_table}\textbf{Growth-based graphs for the prime knots with at most seven crossings.} The first column provides the standard minimal diagram for the studied embedding and the knot's Alexander-Briggs name. The second column shows the idealized graph and valences and corresponding indices of singularities in the associated line field. The third and fourth give the black and white Tait graphs, respectively.} 
\end{figure*}

\subsection{\label{sub:prime}Prime Knots with at most Seven Crossings}

\noindent Figure~\ref{fig:prime_table} summarizes the fourteen nontrivial prime knots with crossing number $c \leq 7$. Within this catalog, our chosen minimal embeddings of the fourteen nontrivial prime knots yield non-isomorphic idealized graphs and defect configurations. The majority of the growth-graphs in Fig.~\ref{fig:prime_table} are three-vertex-connected after idealization, which are isomorphic to graphs formed by vertices and edges of an associated convex polyhedron~\cite{Ziegler1995}. The two-strand torus knots $3_{1}$, $5_{1}$, and $7_{1}$ are an exception to this rule, as they form graphs with two vertices joined by multiple edges. Therefore, all growth-graphs except for torus knots are polyhedral, but all growth-graphs remain three-edge-connected. For the checkerboard convention used in the figure, the growth-based graphs of these torus knots are isomorphic to the white Tait graphs. For the remaining knots, the growth-based graphs are not isomorphic to either Tait graph. They also need not contain either Tait graph as a subgraph. For example, the growth-based graph of the $4_{1}$ knot has more vertices than either Tait graph and does not reproduce their parallel-edge structure.

We should note that reflecting the full 3D initialization leaves the scalar interfacial energy and the label-adjacency rules unchanged. Accordingly, a knot and its mirror produce isomorphic growth-based graphs in our analysis. That is, our construction does not distinguish the handedness of chiral (prime) knots. An oriented spherical embedding could record a reversal of cyclic order, but we leave this for future study.

\subsection{\label{sub:construct}Empirical Reconstruction from Reduced Alternating Diagrams}

\noindent All prime knots with at most seven crossings have reduced alternating diagrams~\cite{Hoste1998TheF1}. This makes it possible to reconstruct the growth-based graphs by a simple diagrammatic rule.

\begin{figure}[!htb]
\includegraphics[width=0.95\columnwidth]{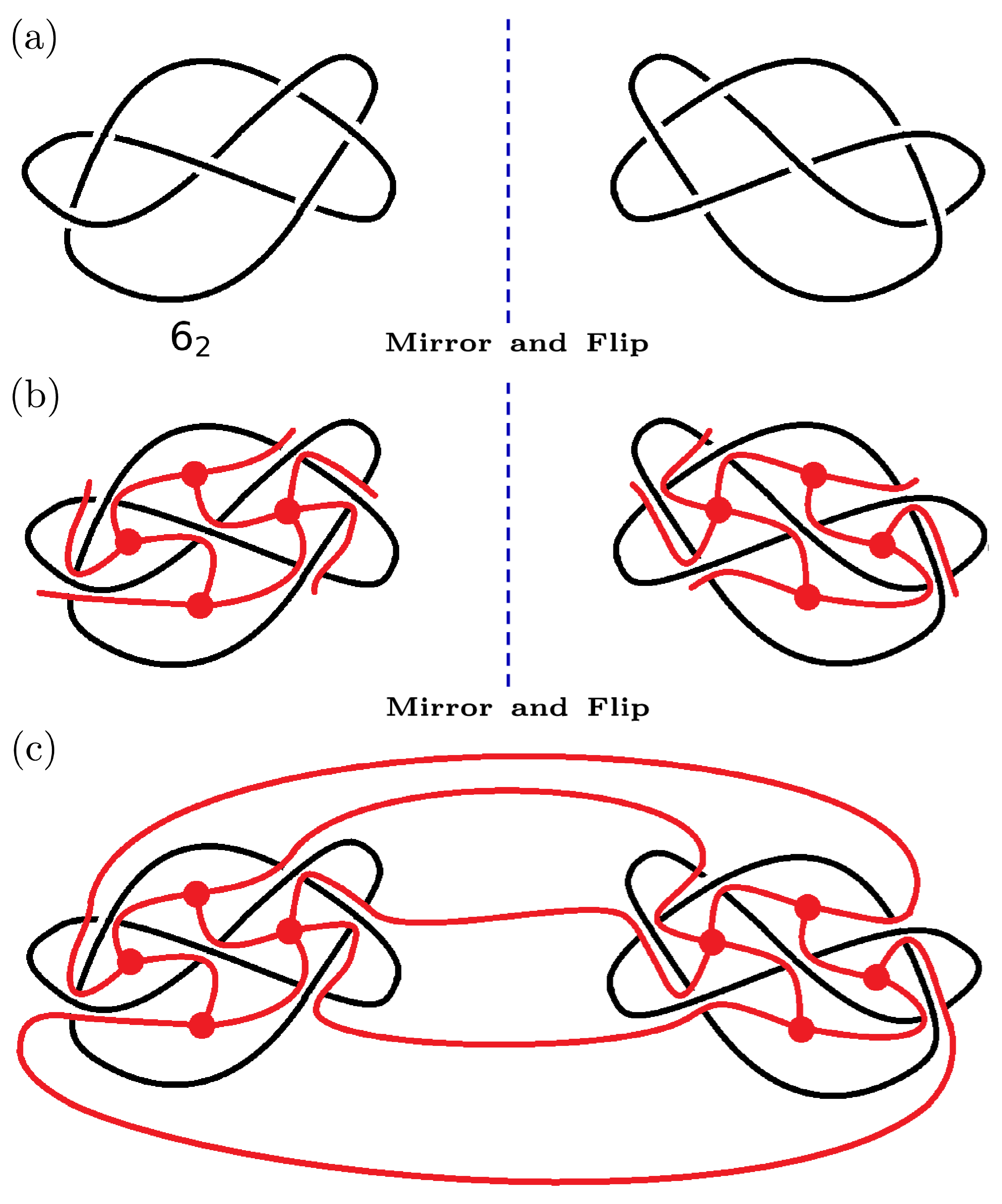}
\caption{\label{fig:algorithm}\textbf{Empirical reconstruction of the growth-based graphs.} (a) Construct the opposite-view diagram, (b) place vertices in tangles according to Fig.~\ref{fig:alternagon}a and extend half-edges through crossings along the under-passing strands, and (c) pair exterior half-edges across the two views in cyclic order. The two diagram interiors represent opposite hemispheres of the final spherical network; the exterior pairings represent equatorial connections.} 
\end{figure}

\begin{figure}[!htb]
\includegraphics[width=1\columnwidth]{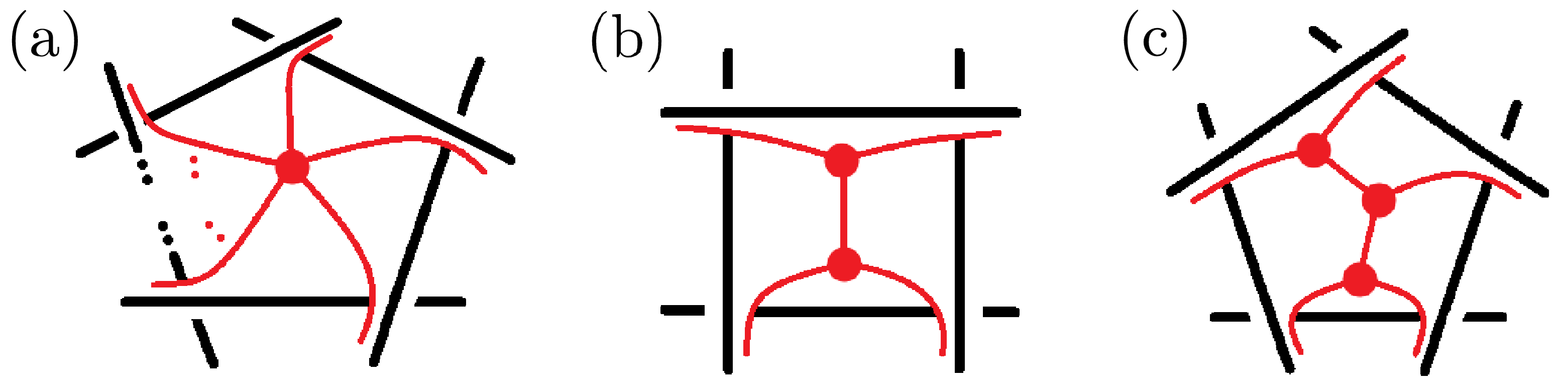}
\caption{\label{fig:alternagon}\textbf{Local templates used in the diagrammatic reconstruction.} (a) An $n$-tangle contributes one valence-$n$ vertex. (b,c) Empirical templates for the non-alternating polygonal regions encountered in the selected connected sums.} 
\end{figure}

Reconstruction proceeds as follows. Let $D$ be an oriented projection of the chosen spatial embedding along a unit vector $\hat{z}$. Then, $D^{\mathrm{opp}}$ is the diagram obtained by viewing the same embedding along $-\hat{z}$, see Fig.~\ref{fig:algorithm}a for an example involving the $6_{2}$ knot~\footnote{We avoid calling $D^{\mathrm{opp}}$ a ``dual diagram,'' because that term is already used for the planar duality of Tait graphs.}. In the drawing convention used here, this operation combines a planar reflection with reversal of every crossing. We refer to a polygonal region of $D$ and $D^{\mathrm{opp}}$ bounded by $n \geq 3$ diagram arcs as a \textit{$n$-tangle} when the over-under roles encountered at successive corners alternate cyclically in the manner that is illustrated in Fig.~\ref{fig:alternagon}.

Next, in every $n$-tangle, place one valence-$n$ vertex and connect it to the $n$ boundary arcs using the rules of Fig.~\ref{fig:alternagon}a. Continue each half-edge through its adjacent crossing along the under-passing strand, as shown in Fig.~\ref{fig:algorithm}b. Finally, pair the half-edges that enter the exterior regions of $D$ and $D^{\mathrm{opp}}$ by their cyclic order around the two projection boundaries, as illustrated in Fig.~\ref{fig:algorithm}c.

The above reconstruction offers a geometric interpretation of the two-part structure visible in the catalog of Fig.~\ref{fig:prime_table}. Arc segments that are overpasses in the projection are exposed preferentially toward the hemisphere viewed along that projection direction, while the corresponding opposite-view overpasses are exposed toward the antipodal hemisphere. The tangle-diagrams collect several comparably exposed arcs forming junction regions. Corresponding edges in the two hemispherical drawings are identified through the projection, thereby connecting the subnetworks into the full spherical graph.

A quantitative derivation of this reconstruction from the interfacial dynamics remains open. However, it is clear that the present reconstruction is not general. The first non-alternating prime knots occur at eight crossings, namely $8_{19}$, $8_{20}$, and $8_{21}$~\cite{Hoste1998TheF1}. Their minimal diagrams contain regions for which the above tangle rule is insufficient. Non-minimal diagrams and connected sums can also contain adjacent non-alternating regions. As we will see in Section~\ref{sub:unknot}, for several non-minimal unknot examples, our reconstruction does not reproduce the expected growth-based graph.

\subsection{\label{sub:select}Selected Connected Sums}

\begin{figure}[!htb]
\includegraphics[width=0.95\columnwidth]{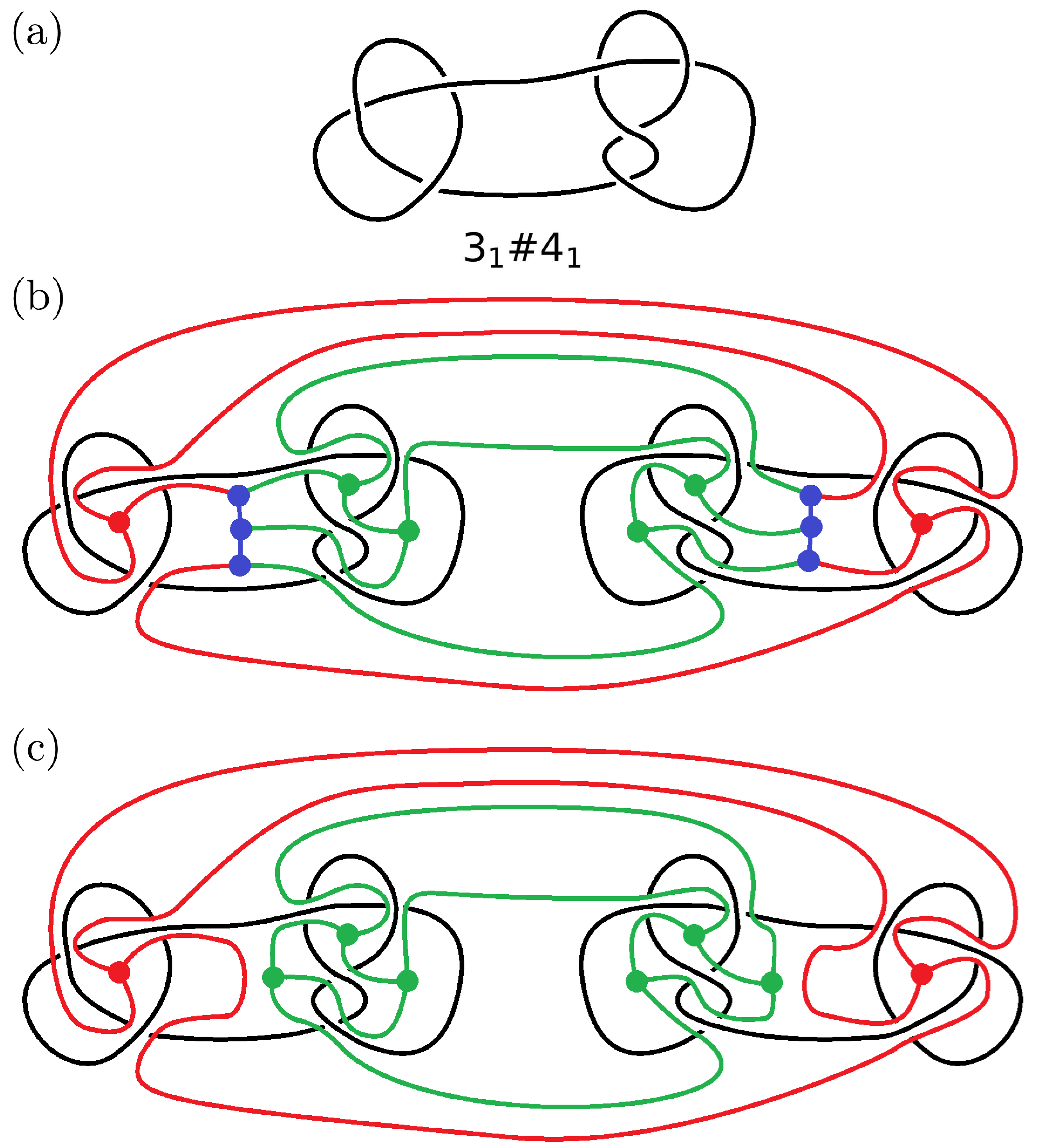}
\caption{\label{fig:composite_algorithm}\textbf{Tracing prime-factor-associated substructures in a $3_{1} \# 4_{1}$ knot sum.} (a) The knot sum under consideration. (b) The composite graph is drawn on the two projection views and partitioned into red and green components associated with the factors, plus blue connecting segments indicating shared parts. (c) To recover one factor, remove the other, recolor the shared edges, retain shared vertices of valence greater than two, and suppress shared vertices of valence two. The recovered graphs agree with the corresponding entries in Fig.~\ref{fig:prime_table} for the $3_{1}$ and $4_{1}$ knot.}
\end{figure}

\noindent We next examine a specific connected sum assembled from a $3_{1}$ and $4_{1}$ knot, see Fig.~\ref{fig:composite_algorithm}a. Our goal is to establish whether features associated with the prime factors remain identifiable in the composite output. We first performed our growth-based analysis, where we retained the cell labels identifying which portion of the initial scaffold originated from each prime factor and propagated these labels to the extracted network. Using the diagrammatic reconstruction of Section~\ref{sub:construct}, the growth-based graph is recovered, see Fig.~\ref{fig:composite_algorithm}b. Here, we use color to distinguish vertices lying entirely inside regions associated with one prime factor. We similarly color edges whose local continuation follows arcs of that factor. The remaining vertices and edges form a shared connector.

To extract the prime-factor-associated portion, we remove the other colored component and absorb shared edges into the retained component. We keep a shared vertex if it has valence greater than two in the retained network, and suppress it if it has valence two. Figure~\ref{fig:composite_algorithm}b,c illustrates the procedure. Note this operation is not a subgraph extraction, because the graph of a prime factor need not occur as a literal subgraph of the sum-knot, growth-based graph.

\begin{figure}[!htb]
\includegraphics[width=1\linewidth]{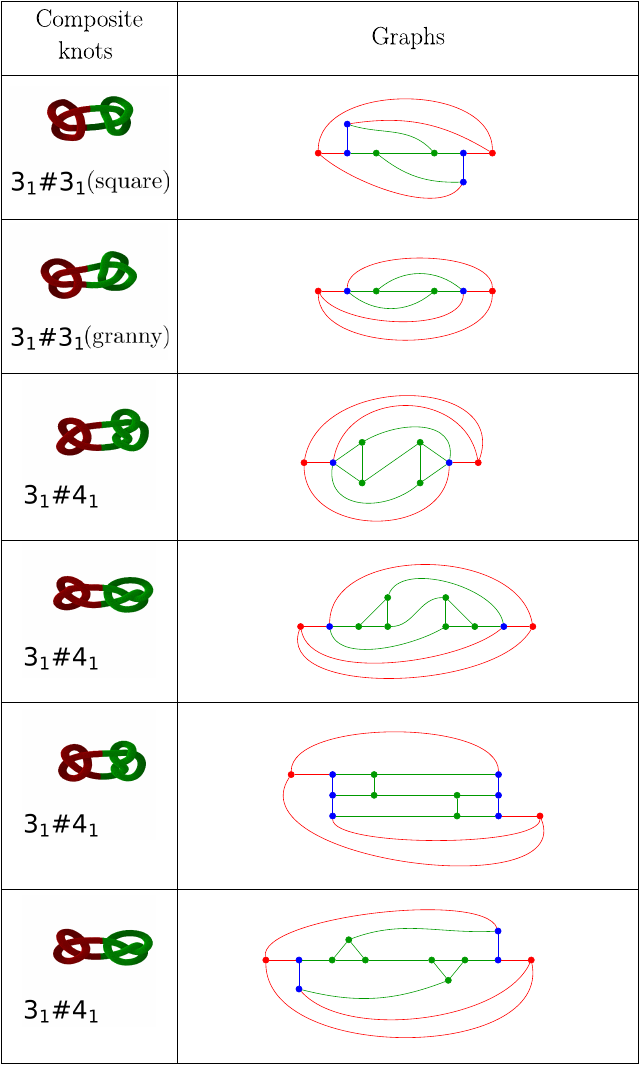}
\caption{\label{tab:tubular}\textbf{Growth-based graphs for selected connected sums.} The diagrams, common name, and Alexander-Briggs notation is given in the left column. The right column shows the growth-based graphs. The red and green identify portions associated with the two prime factors, while blue identifies the shared connector, similar to the coloring in Fig.~\ref{fig:composite_algorithm}.} 
\end{figure}

The square and granny knots provide a useful test of sensitivity to relative chirality~\cite{Hoste1998TheF1}. The square knot is the connected sum of a trefoil with its mirror image and is amphichiral, whereas the granny knot is the connected sum of two trefoils with the same handedness and is chiral. Despite having the same prime factors, the two knots produce distinct growth-based graphs, indicating that the construction is sensitive to the relative handedness of the summands. This is consistent with the mirror blindness observed for prime knots,~\textit{i.e.}, a global reflection reverses the handedness of all chiral factors simultaneously and therefore preserves their relative chirality. We also considered a complementary example in order to sharpen the distinction between sensitivity to relative chirality and invariance under global reflection. The figure-eight knot $4_{1}$ is amphichiral, hence the knots $3_1^{\mathrm{L}}\#4_1$ and $3_1^{\mathrm{R}}\#4_1$ are mirror images of one another. Their growth-based graphs are consequently isomorphic under the present unoriented comparison.

\subsection{\label{sub:unknot}Non-Minimal Diagrams}

\noindent To test invariance under changes of diagram and embedding, we construct ambient-isotopic 3D embeddings for the unknot and selected prime knots, whose chosen projections contain additional Reidemeister-I- or Reidemeister-II-type features. We thus probe the effect of additional local loops and crossing pairs on the growth-based graphs.

\begin{figure*}[!htb]
\includegraphics[width=1\textwidth]{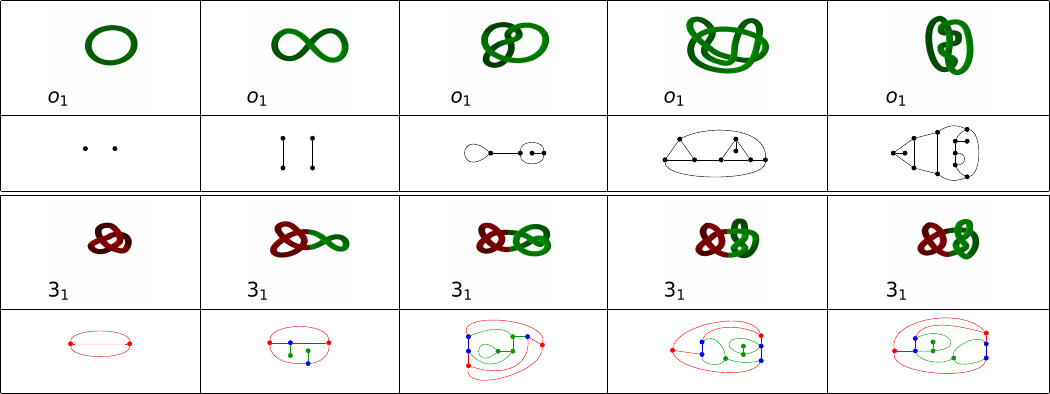}
\caption{\label{fig:homotopy}\textbf{Growth-based graphs for isotoped unknots and trefoils.} Top two rows show the unknot diagrams and the related growth-based graphs, respectively. The bottom two rows repeat the pattern for the trefoil. All examples are related to the minimal knot types (left column) by ambient isotopy. For trefoil examples, red marks the contribution associated with the minimal trefoil and green marks additional unknot-like structure, with blue the connectors as in Fig.~\ref{fig:composite_algorithm}.}
\end{figure*}

Figure~\ref{fig:homotopy} shows non-minimal unknots and trefoils. Equivalent knots can produce non-isomorphic graphs, which directly rules out the raw or idealized graph as a knot invariant. The circular unknot produces a minimal line-field with two $+1$ indices. A local extension and twist can instead produce a graph consisting of two disconnected edges. In our line-field interpretation each endpoint (vertex) carries $+1/2$, though this comes at the cost of not assigning a $+1$ charge to the surface. Presumably this is because the unknot is a degenerate case, and should reasonably be interpreted only in terms of a growth-based graph. The edges connecting the vertices that appear by isotoping the knot can sometimes be removed by applying a contraction.

The same type of graph additions observed for the unknot occur in some non-minimal trefoil diagrams. That is, the original prime-knot contribution remains visible alongside an unknot-like appendage. In these cases, simplifying bridges and suppressing degree-two vertices recovers the minimal idealized graph. Other examples in Fig.~\ref{fig:homotopy} do not admit this simple reduction. Thus far, our analysis has revealed no general graph transformation corresponding to a Reidemeister move.

\begin{figure}[!htb]
\includegraphics[width=1\columnwidth]{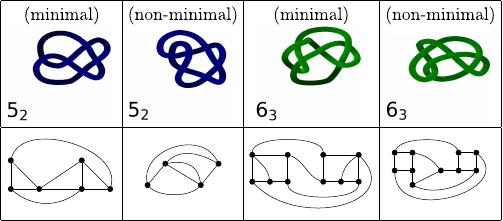}
\caption{\label{fig:homotopy_other}\textbf{Graphs for minimal and non-minimal diagrams of $5_2$ and $6_3$ knots.} The top row shows the diagrams and the bottom row the corresponding growth-based graphs.}
\end{figure}

Lastly, Fig.~\ref{fig:homotopy_other} compares minimal and non-minimal diagrams of the $5_{2}$ and $6_{3}$ knots. These show that some non-minimal diagrams produce graphs with fewer vertices or edges than their minimal counterparts. We conclude that graph size is not monotonic in diagram crossing number. In the displayed $5_{2}$ example, both the number of vertices ($V: 6 \mapsto 4$) and edges ($E: 10 \mapsto 7$) decrease. The $6_{3}$ knot example shows that the number of vertices ($V = 10$) remains unchanged while the edge number ($E: 16 \mapsto 15$) decreases. These results reinforce that the growth process responds to the spatial arrangement; it does not simply relate to the crossing number.

\section{\label{sec:discussion}Discussion}

\noindent Our growth-based calculations demonstrate a reproducible association between knots and graphs for the embeddings that we have considered here. Importantly, these graphs are not generated by projecting the knot and applying a local combinatorial rule, nor by evaluating a conventional knot invariant. Thus, our growth-based graphs should be viewed, at present, as a descriptor of an embedding rather than a canonically assigned object.

Given this limitation, our graph catalog reveals considerably more structure than may be expected from a geometric analysis that is insensitive to the knot type. The protocol produces nonisomorphic idealized graphs. Although this does not imply injectivity, it shows that the surface network retains enough information to discriminate among a nontrivial set of knots. At the same time, the output is not a simple reformulation of a familiar knot-associated graph. Except for the tested two-strand torus knots, for which the result agrees with one member of the checkerboard-based Tait pair, the growth-based graphs differ from both Tait graphs of the corresponding diagrams. However, a systematic comparison with other graph-based representations is required before the growth-based graphs can be claimed to be a new representation of knots and related objects.

Our empirical graph reconstruction helps show that the growth-based graphs that we obtained are strongly constrained by the chosen diagrams. In Section~\ref{sec:results}, we also demonstrated that our rule fails for some non-minimal examples, where local loops and additional crossing pairs alter the graph in ways not captured by our reconstruction approach. Weighting the evidence, a direct diagram-to-graph translation may be possible, but further work is required to identify its properties. This works for a restricted class of embeddings, but cannot at present be elevated to the level of a replacement for the growth calculation. Our non-minimal examples further show that ambient-isotopic representatives of the same knot can produce non-isomorphic graphs. A topological object could still emerge after quotienting by graph transformations induced by ambient isotopy. This would require deriving the changes of the spherical network associated with Reidemeister moves and showing that the resulting graph relations are independent of the chosen move sequence.

Our defect interpretation of the growth-based graph on a sphere attempts to make the connection to the physics of liquid crystals~\cite{nematic_defects, Bowick_2009}. However, it is important to emphasize that our numerical scheme evolves scalar label fields and weighted interfaces. That is, it does not evolve a nematic director or any other microscopic orientational degree of freedom. The line field is instead a completion constructed from the extracted network. A direct physical realization would require an independently defined tangent order parameter whose orientation aligns with the network and whose singularities realize the assigned indices.

This distinction places our approach alongside work on the coupling of geometry, growth, and topology in active and biological matter~\cite{Shankar_2022, BowickSymmetry, giomi2015geometrytopologyturbulenceactive}. In regenerating \textit{Hydra}, for example, topological defects localize at morphological features and can organize mechanical stresses and body-axis formation~\cite{hydra, hernandez2026mechanicsaxisformationtextithydra}. Related defect-mediated mechanisms have been implicated in epithelial remodeling and collective cellular dynamics~\cite{Saw2017TopologicalDI, Vafa_2022, Kawaguchi2016TopologicalDC, hoffmann2022theory}. A complementary biological analogy is plant inosculation, in which growing branches accommodate one another and form persistent contacts without interpenetration~\cite{jefree_yeoman, Yin2012GraftunionDA}. Our construction similarly couples growth and geometric exclusion, but with a global constraint imposed by a knotted scaffold rather than point-like defects. Whether such a knot-to-network conversion can be realized through coupled filament organization, growth, and interfacial mechanics remains an open physical question.

\section{\label{sec:close}Summary and Outlook}

\noindent Summarizing, we have introduced a computational method that maps a 3D knot embedding to a graph on a sphere. This was accomplished by building on the cell-dynamics algorithm introduced by Mohammad~\textit{et al.}~\cite{mohammad2022numerical}. Our approach starts by segmenting a tubular neighborhood of a knot into distinct cells. These are allowed to grow out at a constant rate, while relaxing under a weighted interfacial energy minimization. In the long-time limit, this process leads to an approximately spherical outer boundary, from which we can extract a graph by studying non-neighboring cell contacts. The key element is that the seed region --- the immutable tubular region around the initial segmented knot --- remains connected to the growing cells, allowing the geometry of the initial embedding to influence the resulting surface contact network. The surface graph that is obtained in this manner admits a line-field interpretation with associated topological defects. However, its higher-order junction structure can be sensitive to perturbations, which led us to introduce a junction regularization that contracts selected clusters of nearby threefold junctions into idealized higher-valence vertices,~\textit{e.g.}, in minimal torus knots.

We applied our method to standard minimal-crossing embeddings of the fourteen nontrivial prime knots with at most seven crossings. This led to non-isomorphic graphs, which generally differed from the two Tait graphs of such embeddings. In addition, parts of the introduced growth procedure could be understood in the form of an empirical graph reconstruction. Selected connected sums were shown to retain structures that could be identified as belonging to their prime-knot factors. It should be noted, however, that non-minimal diagrams of the same knot can produce different graphs in our approach. Thus, the growth-based graphs are not knot invariants.

However, our observations provide sufficient clues motivate further investigation in this direction. The open mathematical challenge is to identify graph moves corresponding to Reidemeister moves and to determine whether such a class exists on our diagrams. Our growth-based construction provides a concrete setting in which such graph transformations can be studied.

\section*{\label{sec:ack}Acknowledgments}

\noindent We thank Nikki Fredriksz for analysis of the Mohammad~\textit{et al.}~algorithm~\cite{mohammad2022numerical} that helped motivate the present work. We are also grateful to Boris van Boxtel and Alvaro del Pino Gomez for useful discussions. 

\section*{Data Availability}

\noindent The code, scripts required to reproduce the figures, initial and final embeddings, and supplementary movies are available in an open repository. The persistent identifier is \url{https://doi.org/10.5281/zenodo.24399377}.

\section*{\label{sec:con}Author Contributions}

\noindent Conceptualization, JdG; Methodology, EG \& JdG; Numerical calculations, EG; Validation, EG \& JdG; Investigation, EG; Writing — Original Draft, EG \& JdG; Writing — Review \& Editing, JdG; Resources, JdG; Supervision, JdG. 

\bibliography{knots.bib}

\begin{thebibliography}{47}%
\makeatletter
\providecommand \@ifxundefined [1]{%
 \@ifx{#1\undefined}
}%
\providecommand \@ifnum [1]{%
 \ifnum #1\expandafter \@firstoftwo
 \else \expandafter \@secondoftwo
 \fi
}%
\providecommand \@ifx [1]{%
 \ifx #1\expandafter \@firstoftwo
 \else \expandafter \@secondoftwo
 \fi
}%
\providecommand \natexlab [1]{#1}%
\providecommand \enquote  [1]{``#1''}%
\providecommand \bibnamefont  [1]{#1}%
\providecommand \bibfnamefont [1]{#1}%
\providecommand \citenamefont [1]{#1}%
\providecommand \href@noop [0]{\@secondoftwo}%
\providecommand \href [0]{\begingroup \@sanitize@url \@href}%
\providecommand \@href[1]{\@@startlink{#1}\@@href}%
\providecommand \@@href[1]{\endgroup#1\@@endlink}%
\providecommand \@sanitize@url [0]{\catcode `\\12\catcode `\$12\catcode `\&12\catcode `\#12\catcode `\^12\catcode `\_12\catcode `\%12\relax}%
\providecommand \@@startlink[1]{}%
\providecommand \@@endlink[0]{}%
\providecommand \url  [0]{\begingroup\@sanitize@url \@url }%
\providecommand \@url [1]{\endgroup\@href {#1}{\urlprefix }}%
\providecommand \urlprefix  [0]{URL }%
\providecommand \Eprint [0]{\href }%
\providecommand \doibase [0]{https://doi.org/}%
\providecommand \selectlanguage [0]{\@gobble}%
\providecommand \bibinfo  [0]{\@secondoftwo}%
\providecommand \bibfield  [0]{\@secondoftwo}%
\providecommand \translation [1]{[#1]}%
\providecommand \BibitemOpen [0]{}%
\providecommand \bibitemStop [0]{}%
\providecommand \bibitemNoStop [0]{.\EOS\space}%
\providecommand \EOS [0]{\spacefactor3000\relax}%
\providecommand \BibitemShut  [1]{\csname bibitem#1\endcsname}%
\let\auto@bib@innerbib\@empty
\bibitem [{\citenamefont {Tubiana}\ \emph {et~al.}(2024)\citenamefont {Tubiana}, \citenamefont {Alexander}, \citenamefont {Barbensi}, \citenamefont {Buck}, \citenamefont {Cartwright} \emph {et~al.}}]{TUBIANA20241}%
  \BibitemOpen
  \bibfield  {author} {\bibinfo {author} {\bibfnamefont {L.}~\bibnamefont {Tubiana}}, \bibinfo {author} {\bibfnamefont {G.~P.}\ \bibnamefont {Alexander}}, \bibinfo {author} {\bibfnamefont {A.}~\bibnamefont {Barbensi}}, \bibinfo {author} {\bibfnamefont {D.}~\bibnamefont {Buck}}, \bibinfo {author} {\bibfnamefont {J.~H.~E.}\ \bibnamefont {Cartwright}}, \emph {et~al.},\ }\bibfield  {title} {\bibinfo {title} {Topology in soft and biological matter},\ }\href {https://doi.org/10.1016/j.physrep.2024.04.002} {\bibfield  {journal} {\bibinfo  {journal} {Physics Reports}\ }\textbf {\bibinfo {volume} {1075}},\ \bibinfo {pages} {1} (\bibinfo {year} {2024})}\BibitemShut {NoStop}%
\bibitem [{\citenamefont {Shankar}\ \emph {et~al.}(2022)\citenamefont {Shankar}, \citenamefont {Souslov}, \citenamefont {Bowick}, \citenamefont {Marchetti},\ and\ \citenamefont {Vitelli}}]{Shankar_2022}%
  \BibitemOpen
  \bibfield  {author} {\bibinfo {author} {\bibfnamefont {S.}~\bibnamefont {Shankar}}, \bibinfo {author} {\bibfnamefont {A.}~\bibnamefont {Souslov}}, \bibinfo {author} {\bibfnamefont {M.~J.}\ \bibnamefont {Bowick}}, \bibinfo {author} {\bibfnamefont {M.~C.}\ \bibnamefont {Marchetti}},\ and\ \bibinfo {author} {\bibfnamefont {V.}~\bibnamefont {Vitelli}},\ }\bibfield  {title} {\bibinfo {title} {Topological active matter},\ }\href {https://doi.org/10.1038/s42254-022-00445-3} {\bibfield  {journal} {\bibinfo  {journal} {Nature Reviews Physics}\ }\textbf {\bibinfo {volume} {4}},\ \bibinfo {pages} {380} (\bibinfo {year} {2022})}\BibitemShut {NoStop}%
\bibitem [{\citenamefont {Micheletti}\ \emph {et~al.}(2011)\citenamefont {Micheletti}, \citenamefont {Marenduzzo},\ and\ \citenamefont {Orlandini}}]{Micheletti_2011}%
  \BibitemOpen
  \bibfield  {author} {\bibinfo {author} {\bibfnamefont {C.}~\bibnamefont {Micheletti}}, \bibinfo {author} {\bibfnamefont {D.}~\bibnamefont {Marenduzzo}},\ and\ \bibinfo {author} {\bibfnamefont {E.}~\bibnamefont {Orlandini}},\ }\bibfield  {title} {\bibinfo {title} {Polymers with spatial or topological constraints: Theoretical and computational results},\ }\href {https://doi.org/10.1016/j.physrep.2011.03.003} {\bibfield  {journal} {\bibinfo  {journal} {Physics Reports}\ }\textbf {\bibinfo {volume} {504}},\ \bibinfo {pages} {1} (\bibinfo {year} {2011})}\BibitemShut {NoStop}%
\bibitem [{\citenamefont {Wasserman}\ and\ \citenamefont {Cozzarelli}(1986)}]{Wasserman_DNA}%
  \BibitemOpen
  \bibfield  {author} {\bibinfo {author} {\bibfnamefont {S.~A.}\ \bibnamefont {Wasserman}}\ and\ \bibinfo {author} {\bibfnamefont {N.~R.}\ \bibnamefont {Cozzarelli}},\ }\bibfield  {title} {\bibinfo {title} {Biochemical topology: Applications to {DNA} recombination and replication},\ }\href {https://doi.org/10.1126/science.3010458} {\bibfield  {journal} {\bibinfo  {journal} {Science}\ }\textbf {\bibinfo {volume} {232}},\ \bibinfo {pages} {951} (\bibinfo {year} {1986})}\BibitemShut {NoStop}%
\bibitem [{\citenamefont {Terriac}\ \emph {et~al.}(2019)\citenamefont {Terriac}, \citenamefont {Sch{\"u}tz},\ and\ \citenamefont {Lautenschlaeger}}]{terriac2019vimentin}%
  \BibitemOpen
  \bibfield  {author} {\bibinfo {author} {\bibfnamefont {E.}~\bibnamefont {Terriac}}, \bibinfo {author} {\bibfnamefont {S.}~\bibnamefont {Sch{\"u}tz}},\ and\ \bibinfo {author} {\bibfnamefont {F.}~\bibnamefont {Lautenschlaeger}},\ }\bibfield  {title} {\bibinfo {title} {Vimentin intermediate filament rings deform the nucleus during the first steps of adhesion},\ }\href {https://doi.org/10.3389/fcell.2019.00106} {\bibfield  {journal} {\bibinfo  {journal} {Frontiers in Cell and Developmental Biology}\ }\textbf {\bibinfo {volume} {7}},\ \bibinfo {pages} {106} (\bibinfo {year} {2019})}\BibitemShut {NoStop}%
\bibitem [{\citenamefont {Kleckner}\ and\ \citenamefont {Irvine}(2013)}]{KlecknerIrvine}%
  \BibitemOpen
  \bibfield  {author} {\bibinfo {author} {\bibfnamefont {D.}~\bibnamefont {Kleckner}}\ and\ \bibinfo {author} {\bibfnamefont {W.~T.~M.}\ \bibnamefont {Irvine}},\ }\bibfield  {title} {\bibinfo {title} {Creation and dynamics of knotted vortices},\ }\href {https://doi.org/10.1038/nphys2560} {\bibfield  {journal} {\bibinfo  {journal} {Nature Physics}\ }\textbf {\bibinfo {volume} {9}},\ \bibinfo {pages} {253} (\bibinfo {year} {2013})}\BibitemShut {NoStop}%
\bibitem [{\citenamefont {Raymer}\ and\ \citenamefont {Smith}(2007)}]{Raymer2007}%
  \BibitemOpen
  \bibfield  {author} {\bibinfo {author} {\bibfnamefont {D.~M.}\ \bibnamefont {Raymer}}\ and\ \bibinfo {author} {\bibfnamefont {D.~E.}\ \bibnamefont {Smith}},\ }\bibfield  {title} {\bibinfo {title} {Spontaneous knotting of an agitated string},\ }\href {https://doi.org/10.1073/pnas.0611320104} {\bibfield  {journal} {\bibinfo  {journal} {Proceedings of the National Academy of Sciences}\ }\textbf {\bibinfo {volume} {104}},\ \bibinfo {pages} {16432} (\bibinfo {year} {2007})}\BibitemShut {NoStop}%
\bibitem [{\citenamefont {Patil}\ \emph {et~al.}(2023)\citenamefont {Patil}, \citenamefont {Tuazon}, \citenamefont {Kaufman}, \citenamefont {Chakrabortty}, \citenamefont {Qin}, \citenamefont {Dunkel},\ and\ \citenamefont {Bhamla}}]{Patil_2023}%
  \BibitemOpen
  \bibfield  {author} {\bibinfo {author} {\bibfnamefont {V.~P.}\ \bibnamefont {Patil}}, \bibinfo {author} {\bibfnamefont {H.}~\bibnamefont {Tuazon}}, \bibinfo {author} {\bibfnamefont {E.}~\bibnamefont {Kaufman}}, \bibinfo {author} {\bibfnamefont {T.}~\bibnamefont {Chakrabortty}}, \bibinfo {author} {\bibfnamefont {D.}~\bibnamefont {Qin}}, \bibinfo {author} {\bibfnamefont {J.}~\bibnamefont {Dunkel}},\ and\ \bibinfo {author} {\bibfnamefont {M.~S.}\ \bibnamefont {Bhamla}},\ }\bibfield  {title} {\bibinfo {title} {Ultrafast reversible self-assembly of living tangled matter},\ }\href {https://doi.org/10.1126/science.ade7759} {\bibfield  {journal} {\bibinfo  {journal} {Science}\ }\textbf {\bibinfo {volume} {380}},\ \bibinfo {pages} {392} (\bibinfo {year} {2023})}\BibitemShut {NoStop}%
\bibitem [{\citenamefont {Bowick}\ \emph {et~al.}(2022)\citenamefont {Bowick}, \citenamefont {Fakhri}, \citenamefont {Marchetti},\ and\ \citenamefont {Ramaswamy}}]{BowickSymmetry}%
  \BibitemOpen
  \bibfield  {author} {\bibinfo {author} {\bibfnamefont {M.~J.}\ \bibnamefont {Bowick}}, \bibinfo {author} {\bibfnamefont {N.}~\bibnamefont {Fakhri}}, \bibinfo {author} {\bibfnamefont {M.~C.}\ \bibnamefont {Marchetti}},\ and\ \bibinfo {author} {\bibfnamefont {S.}~\bibnamefont {Ramaswamy}},\ }\bibfield  {title} {\bibinfo {title} {Symmetry, thermodynamics, and topology in active matter},\ }\href {https://doi.org/10.1103/PhysRevX.12.010501} {\bibfield  {journal} {\bibinfo  {journal} {Phys. Rev. X}\ }\textbf {\bibinfo {volume} {12}},\ \bibinfo {pages} {010501} (\bibinfo {year} {2022})}\BibitemShut {NoStop}%
\bibitem [{\citenamefont {Giomi}(2015)}]{giomi2015geometrytopologyturbulenceactive}%
  \BibitemOpen
  \bibfield  {author} {\bibinfo {author} {\bibfnamefont {L.}~\bibnamefont {Giomi}},\ }\bibfield  {title} {\bibinfo {title} {Geometry and topology of turbulence in active nematics},\ }\href {https://doi.org/10.1103/PhysRevX.5.031003} {\bibfield  {journal} {\bibinfo  {journal} {Phys. Rev. X}\ }\textbf {\bibinfo {volume} {5}},\ \bibinfo {pages} {031003} (\bibinfo {year} {2015})}\BibitemShut {NoStop}%
\bibitem [{\citenamefont {Rolfsen}(2003)}]{rolfsen2003knots}%
  \BibitemOpen
  \bibfield  {author} {\bibinfo {author} {\bibfnamefont {D.}~\bibnamefont {Rolfsen}},\ }\href@noop {} {\emph {\bibinfo {title} {Knots and Links}}}\ (\bibinfo  {publisher} {AMS Chelsea Publishing},\ \bibinfo {address} {Providence, RI},\ \bibinfo {year} {2003})\BibitemShut {NoStop}%
\bibitem [{\citenamefont {Lickorish}(1997)}]{LickorishW.B.Raymond1997Aitk}%
  \BibitemOpen
  \bibfield  {author} {\bibinfo {author} {\bibfnamefont {W.~B.~R.}\ \bibnamefont {Lickorish}},\ }\href@noop {} {\emph {\bibinfo {title} {An Introduction to Knot Theory}}},\ \bibinfo {series} {Graduate Texts in Mathematics}, Vol.\ \bibinfo {volume} {175}\ (\bibinfo  {publisher} {Springer},\ \bibinfo {year} {1997})\BibitemShut {NoStop}%
\bibitem [{\citenamefont {Alexander}(1928)}]{Alexander1928TopologicalIO}%
  \BibitemOpen
  \bibfield  {author} {\bibinfo {author} {\bibfnamefont {J.~W.}\ \bibnamefont {Alexander}},\ }\bibfield  {title} {\bibinfo {title} {Topological invariants of knots and links},\ }\href {https://doi.org/10.1090/S0002-9947-1928-1501429-1} {\bibfield  {journal} {\bibinfo  {journal} {Transactions of the American Mathematical Society}\ }\textbf {\bibinfo {volume} {30}},\ \bibinfo {pages} {275} (\bibinfo {year} {1928})}\BibitemShut {NoStop}%
\bibitem [{\citenamefont {Jones}(1985)}]{Jones:1985dw}%
  \BibitemOpen
  \bibfield  {author} {\bibinfo {author} {\bibfnamefont {V.~F.~R.}\ \bibnamefont {Jones}},\ }\bibfield  {title} {\bibinfo {title} {A polynomial invariant for knots via von {Neumann} algebras},\ }\href {https://doi.org/10.1090/S0273-0979-1985-15304-2} {\bibfield  {journal} {\bibinfo  {journal} {Bulletin of the American Mathematical Society}\ }\textbf {\bibinfo {volume} {12}},\ \bibinfo {pages} {103} (\bibinfo {year} {1985})}\BibitemShut {NoStop}%
\bibitem [{\citenamefont {Freyd}\ \emph {et~al.}(1985)\citenamefont {Freyd}, \citenamefont {Yetter}, \citenamefont {Hoste}, \citenamefont {Lickorish}, \citenamefont {Millett},\ and\ \citenamefont {Ocneanu}}]{homfly}%
  \BibitemOpen
  \bibfield  {author} {\bibinfo {author} {\bibfnamefont {P.}~\bibnamefont {Freyd}}, \bibinfo {author} {\bibfnamefont {D.}~\bibnamefont {Yetter}}, \bibinfo {author} {\bibfnamefont {J.}~\bibnamefont {Hoste}}, \bibinfo {author} {\bibfnamefont {W.~B.~R.}\ \bibnamefont {Lickorish}}, \bibinfo {author} {\bibfnamefont {K.}~\bibnamefont {Millett}},\ and\ \bibinfo {author} {\bibfnamefont {A.}~\bibnamefont {Ocneanu}},\ }\bibfield  {title} {\bibinfo {title} {A new polynomial invariant of knots and links},\ }\href {https://doi.org/10.1090/S0273-0979-1985-15361-3} {\bibfield  {journal} {\bibinfo  {journal} {Bulletin of the American Mathematical Society}\ }\textbf {\bibinfo {volume} {12}},\ \bibinfo {pages} {239} (\bibinfo {year} {1985})}\BibitemShut {NoStop}%
\bibitem [{\citenamefont {Przytycki}\ and\ \citenamefont {Traczyk}(1987)}]{homflypt}%
  \BibitemOpen
  \bibfield  {author} {\bibinfo {author} {\bibfnamefont {J.~H.}\ \bibnamefont {Przytycki}}\ and\ \bibinfo {author} {\bibfnamefont {P.}~\bibnamefont {Traczyk}},\ }\bibfield  {title} {\bibinfo {title} {Invariants of links of {Conway} type},\ }\href@noop {} {\bibfield  {journal} {\bibinfo  {journal} {Kobe Journal of Mathematics}\ }\textbf {\bibinfo {volume} {4}},\ \bibinfo {pages} {115} (\bibinfo {year} {1987})}\BibitemShut {NoStop}%
\bibitem [{\citenamefont {Bollob{\'a}s}(1998)}]{Bollobas1998Modern_Ch10}%
  \BibitemOpen
  \bibfield  {author} {\bibinfo {author} {\bibfnamefont {B.}~\bibnamefont {Bollob{\'a}s}},\ }\bibinfo {title} {Modern graph theory}\ (\bibinfo  {publisher} {Springer-Verlag},\ \bibinfo {address} {New York},\ \bibinfo {year} {1998})\ Chap.~\bibinfo {chapter} {10}\BibitemShut {NoStop}%
\bibitem [{\citenamefont {Kauffman}(1989)}]{KAUFFMAN1989105}%
  \BibitemOpen
  \bibfield  {author} {\bibinfo {author} {\bibfnamefont {L.~H.}\ \bibnamefont {Kauffman}},\ }\bibfield  {title} {\bibinfo {title} {A {Tutte} polynomial for signed graphs},\ }\href {https://doi.org/10.1016/0166-218X(89)90049-8} {\bibfield  {journal} {\bibinfo  {journal} {Discrete Applied Mathematics}\ }\textbf {\bibinfo {volume} {25}},\ \bibinfo {pages} {105} (\bibinfo {year} {1989})}\BibitemShut {NoStop}%
\bibitem [{\citenamefont {Thistlethwaite}(1987)}]{THISTLETHWAITE1987297}%
  \BibitemOpen
  \bibfield  {author} {\bibinfo {author} {\bibfnamefont {M.~B.}\ \bibnamefont {Thistlethwaite}},\ }\bibfield  {title} {\bibinfo {title} {A spanning tree expansion of the {Jones} polynomial},\ }\href {https://doi.org/10.1016/0040-9383(87)90003-6} {\bibfield  {journal} {\bibinfo  {journal} {Topology}\ }\textbf {\bibinfo {volume} {26}},\ \bibinfo {pages} {297} (\bibinfo {year} {1987})}\BibitemShut {NoStop}%
\bibitem [{\citenamefont {Bollob{\'a}s}\ and\ \citenamefont {Riordan}(2001)}]{Bollobs2001API}%
  \BibitemOpen
  \bibfield  {author} {\bibinfo {author} {\bibfnamefont {B.}~\bibnamefont {Bollob{\'a}s}}\ and\ \bibinfo {author} {\bibfnamefont {O.}~\bibnamefont {Riordan}},\ }\bibfield  {title} {\bibinfo {title} {A polynomial invariant of graphs on orientable surfaces},\ }\href {https://doi.org/10.1112/plms/83.3.513} {\bibfield  {journal} {\bibinfo  {journal} {Proceedings of the London Mathematical Society}\ }\textbf {\bibinfo {volume} {83}},\ \bibinfo {pages} {513} (\bibinfo {year} {2001})}\BibitemShut {NoStop}%
\bibitem [{\citenamefont {Bollob{\'a}s}\ and\ \citenamefont {Riordan}(2002)}]{Bollobs2002APO}%
  \BibitemOpen
  \bibfield  {author} {\bibinfo {author} {\bibfnamefont {B.}~\bibnamefont {Bollob{\'a}s}}\ and\ \bibinfo {author} {\bibfnamefont {O.}~\bibnamefont {Riordan}},\ }\bibfield  {title} {\bibinfo {title} {A polynomial of graphs on surfaces},\ }\href {https://doi.org/10.1007/s002080100296} {\bibfield  {journal} {\bibinfo  {journal} {Mathematische Annalen}\ }\textbf {\bibinfo {volume} {323}},\ \bibinfo {pages} {81} (\bibinfo {year} {2002})}\BibitemShut {NoStop}%
\bibitem [{\citenamefont {Moffatt}(2008)}]{Moffatt_2008}%
  \BibitemOpen
  \bibfield  {author} {\bibinfo {author} {\bibfnamefont {I.}~\bibnamefont {Moffatt}},\ }\bibfield  {title} {\bibinfo {title} {Knot invariants and the {Bollob{\'a}s--Riordan} polynomial of embedded graphs},\ }\href {https://doi.org/10.1016/j.ejc.2006.12.004} {\bibfield  {journal} {\bibinfo  {journal} {European Journal of Combinatorics}\ }\textbf {\bibinfo {volume} {29}},\ \bibinfo {pages} {95} (\bibinfo {year} {2008})}\BibitemShut {NoStop}%
\bibitem [{\citenamefont {Noble}\ and\ \citenamefont {Welsh}(2000)}]{welsh_knotgraphs}%
  \BibitemOpen
  \bibfield  {author} {\bibinfo {author} {\bibfnamefont {S.~D.}\ \bibnamefont {Noble}}\ and\ \bibinfo {author} {\bibfnamefont {D.~J.~A.}\ \bibnamefont {Welsh}},\ }\bibfield  {title} {\bibinfo {title} {Knot graphs},\ }\href {https://doi.org/10.1002/(SICI)1097-0118(200005)34:1<100::AID-JGT10>3.0.CO;2-9} {\bibfield  {journal} {\bibinfo  {journal} {Journal of Graph Theory}\ }\textbf {\bibinfo {volume} {34}},\ \bibinfo {pages} {100} (\bibinfo {year} {2000})}\BibitemShut {NoStop}%
\bibitem [{\citenamefont {Cantarella}\ \emph {et~al.}(2002)\citenamefont {Cantarella}, \citenamefont {Kusner},\ and\ \citenamefont {Sullivan}}]{Cantarella_2002}%
  \BibitemOpen
  \bibfield  {author} {\bibinfo {author} {\bibfnamefont {J.}~\bibnamefont {Cantarella}}, \bibinfo {author} {\bibfnamefont {R.~B.}\ \bibnamefont {Kusner}},\ and\ \bibinfo {author} {\bibfnamefont {J.~M.}\ \bibnamefont {Sullivan}},\ }\bibfield  {title} {\bibinfo {title} {On the minimum ropelength of knots and links},\ }\href {https://doi.org/10.1007/s00222-002-0234-y} {\bibfield  {journal} {\bibinfo  {journal} {Inventiones Mathematicae}\ }\textbf {\bibinfo {volume} {150}},\ \bibinfo {pages} {257} (\bibinfo {year} {2002})}\BibitemShut {NoStop}%
\bibitem [{\citenamefont {Ashton}\ \emph {et~al.}(2005)\citenamefont {Ashton}, \citenamefont {Cantarella}, \citenamefont {Piatek},\ and\ \citenamefont {Rawdon}}]{ashton2005selfcontactsets50tightly}%
  \BibitemOpen
  \bibfield  {author} {\bibinfo {author} {\bibfnamefont {T.}~\bibnamefont {Ashton}}, \bibinfo {author} {\bibfnamefont {J.}~\bibnamefont {Cantarella}}, \bibinfo {author} {\bibfnamefont {M.}~\bibnamefont {Piatek}},\ and\ \bibinfo {author} {\bibfnamefont {E.~J.}\ \bibnamefont {Rawdon}},\ }\href {https://doi.org/10.48550/arXiv.math/0508248} {\bibinfo {title} {Self-contact sets for 50 tightly knotted and linked tubes}} (\bibinfo {year} {2005}),\ \Eprint {https://arxiv.org/abs/math/0508248} {arXiv:math/0508248 [math.DG]} \BibitemShut {NoStop}%
\bibitem [{\citenamefont {Ashton}\ \emph {et~al.}(2011)\citenamefont {Ashton}, \citenamefont {Cantarella}, \citenamefont {Piatek},\ and\ \citenamefont {Rawdon}}]{Ashton_2011}%
  \BibitemOpen
  \bibfield  {author} {\bibinfo {author} {\bibfnamefont {T.}~\bibnamefont {Ashton}}, \bibinfo {author} {\bibfnamefont {J.}~\bibnamefont {Cantarella}}, \bibinfo {author} {\bibfnamefont {M.}~\bibnamefont {Piatek}},\ and\ \bibinfo {author} {\bibfnamefont {E.~J.}\ \bibnamefont {Rawdon}},\ }\bibfield  {title} {\bibinfo {title} {Knot tightening by constrained gradient descent},\ }\href {https://doi.org/10.1080/10586458.2011.544581} {\bibfield  {journal} {\bibinfo  {journal} {Experimental Mathematics}\ }\textbf {\bibinfo {volume} {20}},\ \bibinfo {pages} {57} (\bibinfo {year} {2011})}\BibitemShut {NoStop}%
\bibitem [{\citenamefont {Mohammad}\ \emph {et~al.}(2022)\citenamefont {Mohammad}, \citenamefont {Murakawa}, \citenamefont {Svadlenka},\ and\ \citenamefont {Togashi}}]{mohammad2022numerical}%
  \BibitemOpen
  \bibfield  {author} {\bibinfo {author} {\bibfnamefont {R.~Z.}\ \bibnamefont {Mohammad}}, \bibinfo {author} {\bibfnamefont {H.}~\bibnamefont {Murakawa}}, \bibinfo {author} {\bibfnamefont {K.}~\bibnamefont {Svadlenka}},\ and\ \bibinfo {author} {\bibfnamefont {H.}~\bibnamefont {Togashi}},\ }\bibfield  {title} {\bibinfo {title} {A numerical algorithm for modeling cellular rearrangements in tissue morphogenesis},\ }\href {https://doi.org/10.1038/s42003-022-03174-6} {\bibfield  {journal} {\bibinfo  {journal} {Communications Biology}\ }\textbf {\bibinfo {volume} {5}},\ \bibinfo {pages} {239} (\bibinfo {year} {2022})}\BibitemShut {NoStop}%
\bibitem [{\citenamefont {Crowley}\ and\ \citenamefont {Grant}(2017)}]{Crowley2016ThePT}%
  \BibitemOpen
  \bibfield  {author} {\bibinfo {author} {\bibfnamefont {D.}~\bibnamefont {Crowley}}\ and\ \bibinfo {author} {\bibfnamefont {M.}~\bibnamefont {Grant}},\ }\bibfield  {title} {\bibinfo {title} {The {Poincar{\'e}-Hopf} theorem for line fields revisited},\ }\href {https://doi.org/10.1016/j.geomphys.2017.03.011} {\bibfield  {journal} {\bibinfo  {journal} {Journal of Geometry and Physics}\ }\textbf {\bibinfo {volume} {117}},\ \bibinfo {pages} {187} (\bibinfo {year} {2017})}\BibitemShut {NoStop}%
\bibitem [{\citenamefont {Shin}\ \emph {et~al.}(2008)\citenamefont {Shin}, \citenamefont {Bowick},\ and\ \citenamefont {Xing}}]{nematic_defects}%
  \BibitemOpen
  \bibfield  {author} {\bibinfo {author} {\bibfnamefont {H.}~\bibnamefont {Shin}}, \bibinfo {author} {\bibfnamefont {M.~J.}\ \bibnamefont {Bowick}},\ and\ \bibinfo {author} {\bibfnamefont {X.}~\bibnamefont {Xing}},\ }\bibfield  {title} {\bibinfo {title} {Topological defects in spherical nematics},\ }\href {https://doi.org/10.1103/PhysRevLett.101.037802} {\bibfield  {journal} {\bibinfo  {journal} {Physical Review Letters}\ }\textbf {\bibinfo {volume} {101}},\ \bibinfo {pages} {037802} (\bibinfo {year} {2008})}\BibitemShut {NoStop}%
\bibitem [{\citenamefont {Bowick}\ and\ \citenamefont {Giomi}(2009)}]{Bowick_2009}%
  \BibitemOpen
  \bibfield  {author} {\bibinfo {author} {\bibfnamefont {M.~J.}\ \bibnamefont {Bowick}}\ and\ \bibinfo {author} {\bibfnamefont {L.}~\bibnamefont {Giomi}},\ }\bibfield  {title} {\bibinfo {title} {Two-dimensional matter: Order, curvature and defects},\ }\href {https://doi.org/10.1080/00018730903043166} {\bibfield  {journal} {\bibinfo  {journal} {Advances in Physics}\ }\textbf {\bibinfo {volume} {58}},\ \bibinfo {pages} {449} (\bibinfo {year} {2009})}\BibitemShut {NoStop}%
\bibitem [{\citenamefont {Merriman}\ \emph {et~al.}(1994)\citenamefont {Merriman}, \citenamefont {Bence},\ and\ \citenamefont {Osher}}]{mbo}%
  \BibitemOpen
  \bibfield  {author} {\bibinfo {author} {\bibfnamefont {B.}~\bibnamefont {Merriman}}, \bibinfo {author} {\bibfnamefont {J.~K.}\ \bibnamefont {Bence}},\ and\ \bibinfo {author} {\bibfnamefont {S.}~\bibnamefont {Osher}},\ }\bibfield  {title} {\bibinfo {title} {Motion of multiple junctions: A level set approach},\ }\href {https://doi.org/10.1006/jcph.1994.1105} {\bibfield  {journal} {\bibinfo  {journal} {Journal of Computational Physics}\ }\textbf {\bibinfo {volume} {112}},\ \bibinfo {pages} {334} (\bibinfo {year} {1994})}\BibitemShut {NoStop}%
\bibitem [{\citenamefont {Esedo{\u{g}}lu}\ and\ \citenamefont {Otto}(2015)}]{esedoglu_otto}%
  \BibitemOpen
  \bibfield  {author} {\bibinfo {author} {\bibfnamefont {S.}~\bibnamefont {Esedo{\u{g}}lu}}\ and\ \bibinfo {author} {\bibfnamefont {F.}~\bibnamefont {Otto}},\ }\bibfield  {title} {\bibinfo {title} {Threshold dynamics for networks with arbitrary surface tensions},\ }\href {https://doi.org/10.1002/cpa.21527} {\bibfield  {journal} {\bibinfo  {journal} {Communications on Pure and Applied Mathematics}\ }\textbf {\bibinfo {volume} {68}},\ \bibinfo {pages} {808} (\bibinfo {year} {2015})}\BibitemShut {NoStop}%
\bibitem [{\citenamefont {Jacobs}\ \emph {et~al.}(2018)\citenamefont {Jacobs}, \citenamefont {Merkurjev},\ and\ \citenamefont {Esedo{\u{g}}lu}}]{jacobs_merkurjev_esedoglu}%
  \BibitemOpen
  \bibfield  {author} {\bibinfo {author} {\bibfnamefont {M.}~\bibnamefont {Jacobs}}, \bibinfo {author} {\bibfnamefont {E.}~\bibnamefont {Merkurjev}},\ and\ \bibinfo {author} {\bibfnamefont {S.}~\bibnamefont {Esedo{\u{g}}lu}},\ }\bibfield  {title} {\bibinfo {title} {Auction dynamics: A volume constrained {MBO} scheme},\ }\href {https://doi.org/10.1016/j.jcp.2017.10.036} {\bibfield  {journal} {\bibinfo  {journal} {Journal of Computational Physics}\ }\textbf {\bibinfo {volume} {354}},\ \bibinfo {pages} {288} (\bibinfo {year} {2018})}\BibitemShut {NoStop}%
\bibitem [{\citenamefont {Plateau}(1873)}]{plateau1873statique}%
  \BibitemOpen
  \bibfield  {author} {\bibinfo {author} {\bibfnamefont {J.}~\bibnamefont {Plateau}},\ }\href@noop {} {\emph {\bibinfo {title} {Statique exp{\'e}rimentale et th{\'e}orique des liquides soumis aux seules forces mol{\'e}culaires}}}\ (\bibinfo  {publisher} {Gauthier-Villars},\ \bibinfo {address} {Paris},\ \bibinfo {year} {1873})\BibitemShut {NoStop}%
\bibitem [{\citenamefont {Taylor}(1976)}]{taylor1976structure}%
  \BibitemOpen
  \bibfield  {author} {\bibinfo {author} {\bibfnamefont {J.~E.}\ \bibnamefont {Taylor}},\ }\bibfield  {title} {\bibinfo {title} {The structure of singularities in soap-bubble-like and soap-film-like minimal surfaces},\ }\href {https://doi.org/10.2307/1970949} {\bibfield  {journal} {\bibinfo  {journal} {The Annals of Mathematics}\ }\textbf {\bibinfo {volume} {103}},\ \bibinfo {pages} {489} (\bibinfo {year} {1976})}\BibitemShut {NoStop}%
\bibitem [{\citenamefont {G{\"u}lmez}\ and\ \citenamefont {de~Graaf}(2026)}]{ElectronicSupplement}%
  \BibitemOpen
  \bibfield  {author} {\bibinfo {author} {\bibfnamefont {E.}~\bibnamefont {G{\"u}lmez}}\ and\ \bibinfo {author} {\bibfnamefont {J.}~\bibnamefont {de~Graaf}},\ }\href@noop {} {\bibinfo {title} {Supplementary information for: ``{Growing Knot Embeddings into Spherical Graphs''}}} (\bibinfo {year} {2026}),\ \bibinfo {note} {available as an ancillary file on arXiv.}\BibitemShut {Stop}%
\bibitem [{\citenamefont {Ziegler}(1995)}]{Ziegler1995}%
  \BibitemOpen
  \bibfield  {author} {\bibinfo {author} {\bibfnamefont {G.~M.}\ \bibnamefont {Ziegler}},\ }\bibinfo {title} {Steinitz' theorem for 3-polytopes},\ in\ \href {https://doi.org/10.1007/978-1-4613-8431-1_4} {\emph {\bibinfo {booktitle} {Lectures on Polytopes: Updated Seventh Printing of the First Edition}}}\ (\bibinfo  {publisher} {Springer New York},\ \bibinfo {address} {New York, NY},\ \bibinfo {year} {1995})\ pp.\ \bibinfo {pages} {103--126}\BibitemShut {NoStop}%
\bibitem [{\citenamefont {Hoste}\ \emph {et~al.}(1998)\citenamefont {Hoste}, \citenamefont {Thistlethwaite},\ and\ \citenamefont {Weeks}}]{Hoste1998TheF1}%
  \BibitemOpen
  \bibfield  {author} {\bibinfo {author} {\bibfnamefont {J.}~\bibnamefont {Hoste}}, \bibinfo {author} {\bibfnamefont {M.}~\bibnamefont {Thistlethwaite}},\ and\ \bibinfo {author} {\bibfnamefont {J.~R.}\ \bibnamefont {Weeks}},\ }\bibfield  {title} {\bibinfo {title} {The first 1,701,936 knots},\ }\href {https://doi.org/10.1007/BF03025227} {\bibfield  {journal} {\bibinfo  {journal} {The Mathematical Intelligencer}\ }\textbf {\bibinfo {volume} {20}},\ \bibinfo {pages} {33} (\bibinfo {year} {1998})}\BibitemShut {NoStop}%
\bibitem [{Note1()}]{Note1}%
  \BibitemOpen
  \bibinfo {note} {We avoid calling $D^{\protect \mathrm {opp}}$ a ``dual diagram,'' because that term is already used for the planar duality of Tait graphs.}\BibitemShut {Stop}%
\bibitem [{\citenamefont {Maroudas-Sacks}\ \emph {et~al.}(2021)\citenamefont {Maroudas-Sacks}, \citenamefont {Garion}, \citenamefont {Shani-Zerbib}, \citenamefont {Livshits}, \citenamefont {Braun},\ and\ \citenamefont {Keren}}]{hydra}%
  \BibitemOpen
  \bibfield  {author} {\bibinfo {author} {\bibfnamefont {Y.}~\bibnamefont {Maroudas-Sacks}}, \bibinfo {author} {\bibfnamefont {L.}~\bibnamefont {Garion}}, \bibinfo {author} {\bibfnamefont {L.}~\bibnamefont {Shani-Zerbib}}, \bibinfo {author} {\bibfnamefont {A.}~\bibnamefont {Livshits}}, \bibinfo {author} {\bibfnamefont {E.}~\bibnamefont {Braun}},\ and\ \bibinfo {author} {\bibfnamefont {K.}~\bibnamefont {Keren}},\ }\bibfield  {title} {\bibinfo {title} {Topological defects in the nematic order of actin fibres as organization centres of {Hydra} morphogenesis},\ }\href {https://doi.org/10.1038/s41567-020-01083-1} {\bibfield  {journal} {\bibinfo  {journal} {Nature Physics}\ }\textbf {\bibinfo {volume} {17}},\ \bibinfo {pages} {251} (\bibinfo {year} {2021})}\BibitemShut {NoStop}%
\bibitem [{\citenamefont {Hernandez}\ \emph {et~al.}(2026)\citenamefont {Hernandez}, \citenamefont {Zhu},\ and\ \citenamefont {Giomi}}]{hernandez2026mechanicsaxisformationtextithydra}%
  \BibitemOpen
  \bibfield  {author} {\bibinfo {author} {\bibfnamefont {A.}~\bibnamefont {Hernandez}}, \bibinfo {author} {\bibfnamefont {C.}~\bibnamefont {Zhu}},\ and\ \bibinfo {author} {\bibfnamefont {L.}~\bibnamefont {Giomi}},\ }\href {https://arxiv.org/abs/2601.05220} {\bibinfo {title} {Mechanics of axis formation in {\textit{hydra}}}} (\bibinfo {year} {2026}),\ \Eprint {https://arxiv.org/abs/2601.05220} {arXiv:2601.05220 [cond-mat.soft]} \BibitemShut {NoStop}%
\bibitem [{\citenamefont {Saw}\ \emph {et~al.}(2017)\citenamefont {Saw}, \citenamefont {Doostmohammadi}, \citenamefont {Nier}, \citenamefont {Kocgozlu}, \citenamefont {Thampi}, \citenamefont {Toyama}, \citenamefont {Marcq}, \citenamefont {Lim}, \citenamefont {Yeomans},\ and\ \citenamefont {Ladoux}}]{Saw2017TopologicalDI}%
  \BibitemOpen
  \bibfield  {author} {\bibinfo {author} {\bibfnamefont {T.~B.}\ \bibnamefont {Saw}}, \bibinfo {author} {\bibfnamefont {A.}~\bibnamefont {Doostmohammadi}}, \bibinfo {author} {\bibfnamefont {V.}~\bibnamefont {Nier}}, \bibinfo {author} {\bibfnamefont {L.}~\bibnamefont {Kocgozlu}}, \bibinfo {author} {\bibfnamefont {S.~P.}\ \bibnamefont {Thampi}}, \bibinfo {author} {\bibfnamefont {Y.}~\bibnamefont {Toyama}}, \bibinfo {author} {\bibfnamefont {P.}~\bibnamefont {Marcq}}, \bibinfo {author} {\bibfnamefont {C.~T.}\ \bibnamefont {Lim}}, \bibinfo {author} {\bibfnamefont {J.~M.}\ \bibnamefont {Yeomans}},\ and\ \bibinfo {author} {\bibfnamefont {B.}~\bibnamefont {Ladoux}},\ }\bibfield  {title} {\bibinfo {title} {Topological defects in epithelia govern cell death and extrusion},\ }\href {https://doi.org/10.1038/nature21718} {\bibfield  {journal} {\bibinfo  {journal} {Nature}\ }\textbf {\bibinfo {volume} {544}},\ \bibinfo {pages} {212} (\bibinfo {year} {2017})}\BibitemShut {NoStop}%
\bibitem [{\citenamefont {Vafa}\ and\ \citenamefont {Mahadevan}(2022)}]{Vafa_2022}%
  \BibitemOpen
  \bibfield  {author} {\bibinfo {author} {\bibfnamefont {F.}~\bibnamefont {Vafa}}\ and\ \bibinfo {author} {\bibfnamefont {L.}~\bibnamefont {Mahadevan}},\ }\bibfield  {title} {\bibinfo {title} {Active nematic defects and epithelial morphogenesis},\ }\href {https://doi.org/10.1103/PhysRevLett.129.098102} {\bibfield  {journal} {\bibinfo  {journal} {Physical Review Letters}\ }\textbf {\bibinfo {volume} {129}},\ \bibinfo {pages} {098102} (\bibinfo {year} {2022})}\BibitemShut {NoStop}%
\bibitem [{\citenamefont {Kawaguchi}\ \emph {et~al.}(2017)\citenamefont {Kawaguchi}, \citenamefont {Kageyama},\ and\ \citenamefont {Sano}}]{Kawaguchi2016TopologicalDC}%
  \BibitemOpen
  \bibfield  {author} {\bibinfo {author} {\bibfnamefont {K.}~\bibnamefont {Kawaguchi}}, \bibinfo {author} {\bibfnamefont {R.}~\bibnamefont {Kageyama}},\ and\ \bibinfo {author} {\bibfnamefont {M.}~\bibnamefont {Sano}},\ }\bibfield  {title} {\bibinfo {title} {Topological defects control collective dynamics in neural progenitor cell cultures},\ }\href {https://doi.org/10.1038/nature22321} {\bibfield  {journal} {\bibinfo  {journal} {Nature}\ }\textbf {\bibinfo {volume} {545}},\ \bibinfo {pages} {327} (\bibinfo {year} {2017})}\BibitemShut {NoStop}%
\bibitem [{\citenamefont {Hoffmann}\ \emph {et~al.}(2022)\citenamefont {Hoffmann}, \citenamefont {Carenza}, \citenamefont {Eckert},\ and\ \citenamefont {Giomi}}]{hoffmann2022theory}%
  \BibitemOpen
  \bibfield  {author} {\bibinfo {author} {\bibfnamefont {L.~A.}\ \bibnamefont {Hoffmann}}, \bibinfo {author} {\bibfnamefont {L.~N.}\ \bibnamefont {Carenza}}, \bibinfo {author} {\bibfnamefont {J.}~\bibnamefont {Eckert}},\ and\ \bibinfo {author} {\bibfnamefont {L.}~\bibnamefont {Giomi}},\ }\bibfield  {title} {\bibinfo {title} {Theory of defect-mediated morphogenesis},\ }\href {https://doi.org/10.1126/sciadv.abk2712} {\bibfield  {journal} {\bibinfo  {journal} {Science Advances}\ }\textbf {\bibinfo {volume} {8}},\ \bibinfo {pages} {eabk2712} (\bibinfo {year} {2022})}\BibitemShut {NoStop}%
\bibitem [{\citenamefont {Jeffree}\ and\ \citenamefont {Yeoman}(1983)}]{jefree_yeoman}%
  \BibitemOpen
  \bibfield  {author} {\bibinfo {author} {\bibfnamefont {C.~E.}\ \bibnamefont {Jeffree}}\ and\ \bibinfo {author} {\bibfnamefont {M.~M.}\ \bibnamefont {Yeoman}},\ }\bibfield  {title} {\bibinfo {title} {Development of intercellular connections between opposing cells in a graft union},\ }\href {https://doi.org/10.1111/j.1469-8137.1983.tb02701.x} {\bibfield  {journal} {\bibinfo  {journal} {New Phytologist}\ }\textbf {\bibinfo {volume} {93}},\ \bibinfo {pages} {491} (\bibinfo {year} {1983})}\BibitemShut {NoStop}%
\bibitem [{\citenamefont {Yin}\ \emph {et~al.}(2012)\citenamefont {Yin}, \citenamefont {Yan}, \citenamefont {Sun}, \citenamefont {Jia}, \citenamefont {Zhang}, \citenamefont {Yan}, \citenamefont {Chai}, \citenamefont {Ren}, \citenamefont {Zheng},\ and\ \citenamefont {Liu}}]{Yin2012GraftunionDA}%
  \BibitemOpen
  \bibfield  {author} {\bibinfo {author} {\bibfnamefont {H.}~\bibnamefont {Yin}}, \bibinfo {author} {\bibfnamefont {B.}~\bibnamefont {Yan}}, \bibinfo {author} {\bibfnamefont {J.}~\bibnamefont {Sun}}, \bibinfo {author} {\bibfnamefont {P.}~\bibnamefont {Jia}}, \bibinfo {author} {\bibfnamefont {Z.}~\bibnamefont {Zhang}}, \bibinfo {author} {\bibfnamefont {X.}~\bibnamefont {Yan}}, \bibinfo {author} {\bibfnamefont {J.}~\bibnamefont {Chai}}, \bibinfo {author} {\bibfnamefont {Z.}~\bibnamefont {Ren}}, \bibinfo {author} {\bibfnamefont {G.}~\bibnamefont {Zheng}},\ and\ \bibinfo {author} {\bibfnamefont {H.}~\bibnamefont {Liu}},\ }\bibfield  {title} {\bibinfo {title} {Graft-union development: A delicate process that involves cell--cell communication between scion and stock for local auxin accumulation},\ }\href {https://doi.org/10.1093/jxb/ers109} {\bibfield  {journal} {\bibinfo  {journal} {Journal of Experimental Botany}\ }\textbf {\bibinfo {volume} {63}},\ \bibinfo {pages} {4219} (\bibinfo {year} {2012})}\BibitemShut
  {NoStop}%
\end{thebibliography}%
 
\end{document}